\documentclass[11pt,a4paper,british,english]{article}
\usepackage[T1]{fontenc}
\usepackage[latin9]{inputenc}
\usepackage{babel}
\usepackage{amsmath}
\usepackage{amssymb}
\usepackage{graphicx}
\usepackage{esint}
\usepackage[unicode=true,pdfusetitle,
 bookmarks=true,bookmarksnumbered=false,bookmarksopen=false,
 breaklinks=false,pdfborder={0 0 1},backref=section,colorlinks=false]
 {hyperref}
\usepackage{breakurl}

\makeatletter

\providecommand{\tabularnewline}{\\}

\numberwithin{equation}{section}
\numberwithin{figure}{section}
\numberwithin{table}{section}

\usepackage{a4wide}\usepackage{psfrag}

\usepackage{babel}

\makeatother

\begin{document}

\title{One-point functions in finite volume/temperature: \\
a case study}

\author{I.M. Szécsényi$^{1}$, G. Takács$^{2,3}$ and G.M.T. Watts$^{4}$\\
 ~\\
 $^{1}$Department of Theoretical Physics, Eötvös University\\
 1117 Budapest, Pázmány Péter sétány 1/A, Hungary\\
 ~\\
 $^{2}$Department of Theoretical Physics, \\
 Budapest University of Technology and Economics\\
 1111 Budapest, Budafoki út 8, Hungary\\
 ~\\
 $^{3}$MTA-BME \textquotedbl{}Momentum\textquotedbl{} Statistical
Field Theory Research Group\\
 1111 Budapest, Budafoki út 8, Hungary\\
 ~\\
 $^{4}$Department of Mathematics, King's College London,\\
 Strand, London WC2R 2LS, UK}

\date{ 4th July 2013}
\maketitle
\begin{abstract}
We consider finite volume (or equivalently, finite temperature) expectation
values of local operators in integrable quantum field theories using
a combination of numerical and analytical approaches. It is shown
that the truncated conformal space approach, when supplemented with
a recently proposed renormalization group, can be sufficiently extended
to the low-energy regime that it can be matched with high precision
by the low-temperature expansion proposed by Leclair and Mussardo.
Besides verifying the consistency of the two descriptions, their combination
leads to an evaluation of expectation values which is valid to a very
high precision for all volume/temperature scales. As a side result
of the investigation, we also discuss some unexpected singularities
in the framework recently proposed by Pozsgay and Takács for the description
of matrix elements of local operators in finite volume, and show that
while some of these singularities are resolved by the inclusion of
the class of exponential finite size corrections known as $\mu$-terms,
these latter corrections themselves lead to the appearance of new
singularities. We point out that a fully consistent description of
finite volume matrix elements is expected to be free of singularities,
and therefore a more complete and systematic understanding of exponential
finite size corrections is necessary. 
\end{abstract}

\section{Introduction}

Finite temperature expectation values of local observables are important
for numerous applications of quantum field theory. In $1+1$ dimensional
integrable quantum field theories, they have been the subject of intensive
studies recently. In \cite{Leclair:1999ys}, Leclair and Mussardo
conjectured a low-temperature expansion which uses two ingredients:
the thermodynamic Bethe Ansatz \cite{Yang:1968rm} as applied to integrable
quantum field theories \cite{Zamolodchikov:1989cf}, and the form
factor bootstrap program \cite{Karowski:1978vz,Kirillov:1987jp,Smirnov:1992vz}.
Although a partial proof of the series was given by Saleur in \cite{Saleur:1999hq},
it was not until the recent introduction of the finite volume form
factor framework by Pozsgay and Takács \cite{Pozsgay:2007kn,Pozsgay:2007gx}
that a systematic construction of the series itself was made possible.
The finite volume form factor formalism even led to an explicit derivation
and further generalization of the series to all orders \cite{Pozsgay:2010xd}.
The series itself has found applications in the investigations of
one-dimensional quantum gases \cite{Kormos:2009zz,Kormos:2010rg,Pozsgay:2011ec}.
The finite volume form factor formalism, on the other hand, was extended
to two-point functions at finite temperature \cite{Essler:2009zz,Essler:2007jp}
and a method to determine the expansion to any given order was developed
in \cite{Pozsgay:2010cr,Szecsenyi:2012jq}.

The aim of the present work is to address some open issues, regarding
both the finite volume form factor formalism and the determination
of thermal one-point functions. Regarding the latter, we make use
of the equivalence between a system at finite temperature $T$ and
in a finite volume $L=1/T$, we consider the one-point function in
finite volume, as determined using the following methods: 
\begin{itemize}
\item the Leclair-Mussardo series; 
\item the thermodynamic Bethe Ansatz (TBA), which is equivalent to the Leclair-Mussardo
(LM) series applied to the trace of the energy-momentum tensor; 
\item and the truncated conformal space approach (TCSA), introduced in \cite{Yurov:1989yu}. 
\end{itemize}
The LM series is a low-energy expansion, valid for large values of
$L$, while the TCSA is a non-perturbative extension of the ultraviolet
perturbation theory, which is valid for small values of $L$. Therefore
care must be taken to find a regime where these approaches can be
matched. This is facilitated by applying a recently introduced TCSA
renormalization group method \cite{Feverati:2006ni,Konik:2007cb,Giokas:2011ix}
to the evaluation of one-point functions, in order to improve the
precision of their TCSA evaluation for larger values of the volume.

In addition to the thermal/finite volume expectations values, this
work also discusses a separate, but related issue that we found when
testing for the consistency of the TCSA and the bootstrap form factors.
Namely, due to singular \foreignlanguage{british}{behaviour} in the
density of states, the finite volume form factors predicted by the
formulae in \cite{Pozsgay:2007kn,Pozsgay:2007gx} show some singularities.
It turns out that these can be resolved by including some of the exponential
finite size corrections, the so-called $\mu$-terms along the lines
of \cite{Pozsgay:2008bf}.

For the most detailed calculations we chose the $T_{2}$ model which
is an integrable perturbation of the minimal conformal field theory
$\mathcal{M}_{2,7}$ by its relevant operator $\Phi_{1,3}$. This
was motivated by the fact that this is one of the simplest theories
in which there is an additional non-trivial primary field besides
the perturbing one, so this provides a non-trivial application of
the LM series in the sense that the series for this operator is not
a consequence of the TBA. In addition, we also study numerically the
scaling Lee-Yang model, and consider the implications of our results
for other models that were treated by TCSA elsewhere in the literature. 

The outline of the paper is as follows. In section \ref{sec:One-point-functions}
we briefly recall relevant facts about the LM series and TBA. In section
\ref{sec:One-point-functions-from-TCSA-RG} we give a derivation of
the cut-off dependence of expectation values in TCSA, since we need
the explicit exponents for our calculations. While the extrapolation
has already been used on at least two occasions \cite{Konik:2007cb,Palmai:2012kf},
the theory behind the exponents has never been exposed and our derivation
fills this gap. Section \ref{sec:Numerical-comparison} is devoted
to the comparison of the one-point function obtained from the TCSA
on the one hand, and the LM series (or, equivalently, from the TBA
in the case of the trace of the energy-momentum tensor) on the other.
In section \ref{sec:A-remark-on-finite-volume-form-factors-and-mu-terms}
we digress a little to discuss the issue of singularities in the finite
volume form factor formalism and their resolution by the inclusion
of $\mu$-term corrections. Section \ref{sec:Conclusions-and-outlook}
is reserved for the conclusions\@. There are two appendices: Appendix
\ref{sec:TCSA-in-T2} gives the relevant details about TCSA and its
application to the $T_{2}$ model, while Appendix \ref{sec:Scattering-theory-of-T2}
is a brief summary of its scattering theory and form factors.

\section{One-point functions at finite temperature \label{sec:One-point-functions} }

\subsection{Perturbed conformal field theories}

To set up notations, we define the class of models we consider with
the formal (Minkowskian) action
\begin{equation}
\mathcal{A}=\mathcal{A}_{CFT}-\lambda\int dtdxV(t,x)\label{eq:PCFT_Maction}
\end{equation}
where $\mathcal{A}_{CFT}$ is a $1+1$ dimensional conformal field
theory corresponding to the ultraviolet fixed point, and $V$ is a
relevant field with zero spin, i.e. with equal left and right conformal
weights $h_{V}=\bar{h}_{V}<1$. Such a theory has a Hamiltonian of
the form
\begin{equation}
H=H_{*}+\lambda\int dxV\label{eq:PCFT_Hamiltonian}
\end{equation}
where $H_{*}$ is the Hamiltonian of the conformal fixed point theory.
The coupling $\lambda$ is dimensionful and can be related to the
physical mass scale $m$ characterizing the off-critical dynamics
by a relation of the form
\begin{equation}
\lambda=\kappa m^{2-2h_{V}}
\end{equation}
with $\kappa$ a dimensionless constant. In the examples considered
here, we identify $m$ with the mass of the lightest particle, i.e.
the gap between the lowest excitation and the vacuum (in infinite
volume).

For most of the paper we consider the $T_{2}$ model, which is defined
as the perturbation of the $\mathcal{M}_{2,7}$ minimal model by its
$\Phi_{1,3}$ operator. The model itself is described in Appendices
\ref{sec:TCSA-in-T2} and \ref{sec:Scattering-theory-of-T2}, where
relevant data regarding the conformal field theory, the exact infinite
volume expectation values and the factorized scattering theory can
be found.

\subsection{The Leclair-Mussardo series\label{sub:The-Leclair-Mussardo-series}}

\subsubsection{The LM series itself\label{sub:The-LM-series}}

The conjecture for the finite volume expectation value of a local
operator by Leclair and Mussardo \cite{Leclair:1999ys} takes the
following form in the $T_{2}$ model:

\begin{eqnarray}
\left\langle \mathcal{O}\right\rangle _{L} & = & \sum_{n,m=0}^{\infty}\frac{1}{n!m!}\int_{-\infty}^{\infty}\frac{\mathrm{d}\theta_{1}}{2\pi}\dots\frac{\mathrm{d}\theta_{n}}{2\pi}\frac{\mathrm{d}\tilde{\theta}_{1}}{2\pi}\dots\frac{\mathrm{d}\tilde{\theta}_{n}}{2\pi}\prod_{i=1}^{n}\frac{1}{1+e^{\varepsilon_{1}\left(\theta_{i}\right)}}\prod_{j=1}^{m}\frac{1}{1+e^{\varepsilon_{2}\left(\tilde{\theta}_{j}\right)}}\nonumber \\
 &  & \times F_{2n,2m,c}^{\mathcal{O}}\left(\theta_{1},\dots,\theta_{n},\tilde{\theta}_{1},\dots,\tilde{\theta}_{m}\right)\label{eq:LMseries}
\end{eqnarray}
where $\theta_{i}$, $\tilde{\theta}_{i}$ are rapidities of the particles
with mass $m_{1}$ and $m_{2}$. The $\varepsilon_{a}$ are the pseudo-energy
functions that satisfy the TBA integral equation 
\begin{eqnarray}
\varepsilon_{a}\left(\theta\right) & = & m_{a}L\cosh\theta-\sum_{b}\int_{-\infty}^{\infty}\frac{\mathrm{d}\theta'}{2\pi}\varphi_{ab}\left(\theta-\theta'\right)\log\left(1+e^{-\varepsilon_{b}\left(\theta'\right)}\right)\label{T2TBA}
\end{eqnarray}
with $\varphi_{ab}$ the derivatives of the phase shift of the two-particle
S-matrices:
\begin{equation}
\varphi_{ab}(\theta)=-i\frac{d\log S_{ab}(\theta)}{d\theta}
\end{equation}
 and $F_{2n,2m,c}^{\mathcal{O}}$ is the connected form factor defined
by 
\begin{eqnarray}
 &  & F_{2n,2m,c}^{\mathcal{O}}\left(\theta_{1},\dots,\theta_{n},\tilde{\theta}_{1},\dots,\tilde{\theta}_{m}\right)=\nonumber \\
 &  & \mathrm{finite\: part\: of\:}F_{\underbrace{{\scriptstyle 2\dots2}}_{m}\underbrace{{\scriptstyle 1\dots1}}_{n}\underbrace{{\scriptstyle 1\dots1}}_{n}\underbrace{{\scriptstyle 2\dots2}}_{m}}^{\mathcal{O}}\Big(\tilde{\theta}_{m}+i\pi+\epsilon_{m+n},\dots,\tilde{\theta}_{1}+i\pi+\epsilon_{m+1},\nonumber \\
 &  & \theta_{n}+i\pi+\epsilon_{n+1},\dots,\theta_{1}+i\pi+\epsilon_{1},\theta_{1},\dots,\theta_{n},\tilde{\theta}_{1},\dots,\tilde{\theta}_{m}\Big)\label{eq:connectedFF}
\end{eqnarray}
where the finite part is taken by expanding the $\epsilon$-dependence
of the form factor expression for small values of the $\epsilon$-s
and omitting all terms which blow up when any of the $\epsilon_{k}$
is taken to $0$ (including terms that depend on the ratios of the
$\epsilon_{k}$) \cite{Leclair:1999ys}.

\subsubsection{Evidence for the LM series: TBA and finite volume form factors\label{sub:Evidence-for-the}}

Leclair and Mussardo showed that for the trace of energy momentum
tensor their series coincides with the result derived from the TBA
by Zamolodchikov \cite{Zamolodchikov:1989cf}, which we briefly summarize
below specialized to the $T_{2}$ model.

Since $T_{2}$ is the perturbation of $\mathcal{M}_{2,7}$ minimal
model by $\Phi_{1,3}$ operator, the trace of the stress energy tensor
is given by%
\footnote{The stress energy tensor is normalized following the conventions of
Zamolodchikov's paper \cite{Zamolodchikov:1989cf}.%
}

\begin{eqnarray}
\Theta & = & T_{\mu}^{\mu}=2\pi\lambda(2h_{1,3}-2)\Phi_{1,3}\label{eq:Theta_Phi13}
\end{eqnarray}
where $\lambda$ is the coupling constant and $h_{1,3}$ the conformal
weight of the operator. The ground state energy in finite volume is
given by

\begin{equation}
E_{0}(L)=-\sum_{a}\int_{-\infty}^{\infty}\frac{d\theta}{2\pi}m_{a}\cosh\theta\log\left(1+e^{-\varepsilon_{a}\left(\theta\right)}\right)
\end{equation}
where the $\varepsilon_{a}$ solve (\ref{T2TBA}). The vacuum expectation
value of $\Theta$ can be expressed by the following derivative of
$E_{0}\left(L\right)$ 
\begin{eqnarray}
\left\langle \Theta\right\rangle _{L} & = & 2\pi\frac{1}{L}\frac{d}{dL}\left[LE_{0}\left(L\right)\right]=2\pi\left[\frac{E_{0}\left(L\right)}{L}+\frac{dE_{0}\left(L\right)}{dL}\right]\nonumber \\
 & = & \sum_{a}m_{a}\int_{-\infty}^{\infty}d\theta\frac{1}{1+e^{\varepsilon_{a}\left(\theta\right)}}\left[\cosh\theta\partial_{L}\varepsilon_{a}\left(\theta\right)-\frac{1}{L}\sinh\theta\partial_{\theta}\varepsilon_{a}\left(\theta\right)\right]\label{eq:thetaTBA}
\end{eqnarray}
where the derivatives of the TBA pseudo-energy solve the following
linear equations

\begin{eqnarray}
\partial_{L}\varepsilon_{a}\left(\theta\right)=m_{a}\cosh\theta+\sum_{b}\int_{-\infty}^{\infty}\frac{d\theta'}{2\pi}\varphi_{ab}\left(\theta-\theta'\right)\frac{1}{1+e^{\varepsilon_{b}\left(\theta'\right)}}\partial_{L}\varepsilon_{b}\left(\theta'\right)\nonumber \\
\partial_{\theta}\varepsilon_{a}\left(\theta\right)=m_{a}L\sinh\theta+\sum_{b}\int_{-\infty}^{\infty}\frac{d\theta'}{2\pi}\varphi_{ab}\left(\theta-\theta'\right)\frac{1}{1+e^{\varepsilon_{b}\left(\theta'\right)}}\partial_{\theta}\varepsilon_{b}\left(\theta'\right)\label{eq:pseudoenderTBA}
\end{eqnarray}
By expanding the solution to these equations for large $L$, one obtains
the series (\ref{eq:LMseries}) \cite{Leclair:1999ys}. Extending
the argument, Saleur proved the equivalence of the TBA and the LM
series for one-point functions of conserved charge densities \cite{Saleur:1999hq}.

For general operators where a TBA construction is not known, the LM
series was demonstrated to be true up to (and including) three particle
terms in \cite{Pozsgay:2007gx}, and a proof to all orders was presented
in \cite{Pozsgay:2010xd}. We remark that these results were proven
using the finite volume form factor formalism introduced in \cite{Pozsgay:2007kn,Pozsgay:2007gx},
and the proofs relies on the expression of finite volume diagonal
matrix elements (up to exponential corrections in the volume), which
is still only a conjecture, albeit well-supported both by analytical
arguments \cite{Saleur:1999hq,Pozsgay:2007gx} and numerical data
\cite{Pozsgay:2007gx}.

\section{One-point functions from TCSA using RG extrapolation \label{sec:One-point-functions-from-TCSA-RG}}

\subsection{Cut-off dependence of the vacuum expectation value}

For a numerical evaluation of the expectation values, we used the
truncated conformal space approach (TCSA) developed by Yurov and Zamolodchikov
in \cite{Yurov:1989yu}; for the technical details we refer the reader
to appendix \ref{sec:TCSA-in-T2}.

In order to test the exponentially decaying finite volume corrections
predicted by the LM series, we need to improve the accuracy of the
TCSA. This can be achieved by extrapolating in the cut-off (truncation
level) dependence using TCSA RG, which was first applied in \cite{Konik:2007cb},
and later also in \cite{Palmai:2012kf}. However, the way to derive
the necessary exponents has not yet been presented; therefore here
we give the necessary details below so that we can apply it to our
case, and also in order to facilitate future applications of the method.

The derivation of the cut-off dependence below uses the general ideas
presented in \cite{Giokas:2011ix} where a similar method was used
to obtain the running coupling constants. Using the fact that we are
dealing with a relevant perturbation, the high energy \foreignlanguage{british}{behaviour}
related to the cutoff is perturbative in nature, and we perform perturbation
theory in the Euclidean formalism where the action takes the form
\begin{equation}
\mathcal{A}^{E}=\mathcal{A}_{CFT}^{E}+\lambda\int d\tau dxV\label{eq:PCFT_Eaction}
\end{equation}
where $\tau$ is Euclidean time.

We parametrize the cut-off $e_{\mathrm{cut}}$ by the left descendant
quantum number of the highest vectors included in the truncated space.
In the $T_{2}$ model, this is the same for all modules in the Hilbert
space as all the differences between the primary dimensions are smaller
than $2$. The vacuum expectation value of the operator $\mathcal{O}$,
when cutting TCSA at level $n$ is: 
\begin{equation}
E_{\mathcal{O}}(n)=\frac{Q(n)}{\mathcal{N}(n)}
\end{equation}

\begin{equation}
Q(n)=\left\langle \mbox{min}\right|P_{n}\mathcal{O}(0,0)P_{n}e^{-\lambda\int d^{2}zP_{n}V(z,\bar{z})P_{n}}|\mbox{min}\rangle
\end{equation}
where $z=\tau-ix$, $V$ is the perturbing operator, which has conformal
spin zero (i.e. $h_{V}=\bar{h}_{V}$) to ensure translational invariance;
to simplify the exposition, we also assume that $h_{\mathcal{O}}=\bar{h}_{\mathcal{O}}$.
All operator products are assumed to be time ordered, $P_{n}$ is
the projector on states with descendant numbers less than or equal
to $n$ and $|\mbox{min}\rangle$ is the lowest weight primary state,
whose perturbation becomes the vacuum of the massive theory (for $T_{2}$
this is the state created by $\Phi_{1,3}$). The normalization factor
is 
\begin{equation}
\mathcal{N}(n)=\left\langle \mbox{min}\right|e^{-\lambda\int d^{2}zP_{n}V(z,\bar{z})P_{n}}|\mbox{min}\rangle
\end{equation}
Our aim is to calculate the dependence on $n$, i.e. the difference
$Q(n)-Q(n-1)$. To simplify the subsequent formulae, let us introduce
the following notation for the past and future evolution operators
corresponding to the cut-off Hamiltonian: 
\begin{equation}
U_{\pm}^{(n)}(\lambda)=T\exp\left(-\lambda\int_{\tau'\gtrless0}d^{2}z'P_{n}V(z',\bar{z}')P_{n}\right)
\end{equation}
so we can make time ordering explicit 
\begin{equation}
Q(n)=\left\langle \mbox{min}\right|U_{+}^{(n)}(\lambda)P_{n}\mathcal{O}(0,0)P_{n}U_{-}^{(n)}(\lambda)|\mbox{min}\rangle
\end{equation}
We assume that $n$ is large, and so the states at descendant level
$n$ have energy much higher than the mass gap $m$: 
\begin{equation}
\frac{4\pi n}{L}\gg m\label{eq:CPTcondition}
\end{equation}
When this condition is satisfied, we can evaluate their contribution
to first order in $\lambda$. Defining the projector on the level
$n$ descendants as $\tilde{P}_{n}=P_{n}-P_{n-1}$ and denoting 
\begin{equation}
\Delta_{n}V(z,\bar{z})=P_{n-1}V(z,\bar{z})\tilde{P}_{n}+\tilde{P}_{n}V(z,\bar{z})P_{n-1}+\tilde{P}_{n}V(z,\bar{z})\tilde{P}_{n}
\end{equation}
the $\tau<0$ time-evolution operator can be expanded as 
\begin{eqnarray}
U_{-}^{(n)}(\lambda) & = & T\exp\left(-\lambda\int_{\tau<0}d^{2}z\left(P_{n-1}V(z,\bar{z})P_{n-1}+\Delta_{n}V(z,\bar{z})\right)\right)=\\
 &  & T\bigg\{\left[1-\lambda\int_{\tau<0}d^{2}z\Delta_{n}V(z,\bar{z})\right]\exp\left(-\lambda\int_{\tau'<0}d^{2}z'P_{n-1}V(z',\bar{z}')P_{n-1}\right)\bigg\}+O(\lambda^{2})\nonumber \\
 & = & \left[1-\lambda\int_{\tau<0}d^{2}z\Delta_{n}V(z,\bar{z})\right]U_{-}^{(n-1)}(\lambda)+O(\lambda^{2})
\end{eqnarray}
and similarly for $U_{+}^{(n)}(\lambda)$. This yields 
\begin{eqnarray}
Q(n)-Q(n-1) & = & \left\langle \mbox{min}\right|\,\Bigg\{ U_{+}^{(n-1)}(\lambda)\left[1-\lambda\int_{\tau>0}d^{2}zV(z,\bar{z})\tilde{P}_{n}\right]\left(P_{n-1}+\tilde{P}_{n}\right)\nonumber \\
 &  & \quad\times\mathcal{O}(0,0)\left(P_{n-1}+\tilde{P}_{n}\right)\left[1-\lambda\int_{\tau<0}d^{2}z\tilde{P}_{n}V(z,\bar{z})\right]U_{-}^{(n-1)}(\lambda)\Bigg\}\,\left|\mbox{min}\right\rangle \nonumber \\
 &  & -\left\langle \mbox{min}\right|U_{+}^{(n-1)}(\lambda)P_{n-1}\mathcal{O}(0,0)P_{n-1}U_{-}^{(n-1)}(\lambda)|\mbox{min}\rangle+O(\lambda^{2})
\end{eqnarray}
where we used that as a consequence of 
\begin{equation}
P_{n-1}\tilde{P}_{n}=\tilde{P}_{n}P_{n-1}=0
\end{equation}
$\tilde{P}_{n}$ commutes with $U_{\pm}^{(n-1)}(\lambda)$, and also
the relations 
\begin{eqnarray}
\tilde{P}_{n}\left|\mbox{min}\right\rangle  & = & 0\nonumber \\
P_{n-1}\left|\mbox{min}\right\rangle  & = & \left|\mbox{min}\right\rangle 
\end{eqnarray}
Exploiting invariance under time reflection $\tau\rightarrow-\tau$
\begin{eqnarray}
Q(n)-Q(n-1) & = & -2\lambda\int_{\tau<0}d^{2}z\,\left\langle \mbox{min}\right|U_{+}^{(n-1)}(\lambda)\,\mathcal{O}(0,0)\tilde{P}_{n}V(z,\bar{z})\, U_{-}^{(n-1)}(\lambda)\left|\mbox{min}\right\rangle \nonumber \\
 &  & +O(\lambda^{2})\label{eqn:deltaq1}
\end{eqnarray}
Applying the exponential map (\ref{eq:exponentialmap}) and writing
$w=re^{-i\varphi}$ 
\begin{equation}
\int_{\tau<0}d^{2}z=\left(\frac{L}{2\pi}\right)^{2}\int_{0}^{1}\frac{dr}{r}\int_{0}^{2\pi}d\varphi
\end{equation}
yields 
\begin{eqnarray}
 &  & Q(n)-Q(n-1)=\nonumber \\
 &  & -2\lambda\left(\frac{2\pi}{L}\right)^{2h_{\mathcal{O}}+2h_{V}-2}\int_{0}^{1}\frac{dr}{r}\int_{0}^{2\pi}d\varphi\, r^{2h_{V}}\left\langle \mbox{min}\right|U_{+}^{(\infty)}(\lambda)\,\mathcal{O}(1,1)\tilde{P}_{n}V(w,\bar{w})\, U_{-}^{(\infty)}(\lambda)\left|\mbox{min}\right\rangle \nonumber \\
 &  & +O(\lambda^{2})\label{eq:diffQ}
\end{eqnarray}
where we used that to leading order in the coupling $\lambda$ one
can take $n\rightarrow\infty$ in the time evolution exponentials,
since from our calculations it is clear that their cut-off dependence
is itself of order $\lambda$. Using the operator product expansion
(OPE) in conformal field theory 
\begin{equation}
\mathcal{O}(1,1)V(w,\bar{w})=\sum_{A}\frac{C_{\mathcal{O}V}^{A}\, A(1,1)}{(1-w)^{h_{\mathcal{O}}+h_{V}-h_{A}}(1-\bar{w})^{\bar{h}_{\mathcal{O}}+\bar{h}_{V}-\bar{h}_{A}}}\label{eq:twoptOPE}
\end{equation}
where $A$ runs over the set of scaling fields and the $C$ are the
operator product structure constants. For the minimal models we consider
later, the detailed structure of the OPE together with the structure
constants can be found in \cite{Dotsenko:1984nm,Dotsenko:1984ad,Dotsenko:1985hi}.

As described in Appendix \ref{sec:TCSA-in-T2}, due to the fact that
the perturbing operator satisfied $h_{V}=\bar{h}_{V}$, the TCSA Hilbert
space decomposes into sectors with a given value of the conformal
spin $L_{0}-\bar{L}_{0}$, which is in fact a consequence of translational
invariance. Since the vacuum state is translationally invariant it
has zero conformal spin. Therefore all intermediate states contributing
to the matrix element have spin zero, and so the operators $A$ contributing
to the cutoff dependence also have zero conformal spin, i.e. $h_{A}=\bar{h}_{A}$
as well. Substituting (\ref{eq:twoptOPE}) into (\ref{eq:diffQ})
and using the expansions

\begin{eqnarray}
\frac{1}{(1-w)^{\alpha}} & = & \sum_{n=0}^{\infty}\frac{\Gamma(\alpha+n)}{\Gamma(\alpha)\Gamma(n+1)}w^{n}\nonumber \\
\frac{1}{(1-\bar{w})^{\alpha}} & = & \sum_{\bar{n}=0}^{\infty}\frac{\Gamma(\alpha+\bar{n})}{\Gamma(\alpha)\Gamma(\bar{n}+1)}\bar{w}^{\bar{n}}
\end{eqnarray}
the $(n,\bar{n})$ term just gives the contribution to the $\mathcal{O}V$
two-point function from intermediate states at level $(n,\bar{n})$,
which corresponds to the projector $\tilde{P}_{n}$ in eqn. (\ref{eq:diffQ}).
Due to the angular integral the only contributing terms are those
in which $\bar{n}=n$, which is another manifestation of translational
invariance. The exponentials ensure that the contribution of a given
operator $A$ in the OPE will be proportional to the exact finite
volume vacuum expectation value in the perturbed theory 
\begin{equation}
\left.\left\langle \mbox{min}\right|U_{+}^{(\infty)}(\lambda)A(w,\bar{w})U_{-}^{(\infty)}(\lambda)\left|\mbox{min}\right\rangle \right|_{w=\bar{w}=1}
\end{equation}
This captures the non-perturbative infrared physics, while perturbation
theory applies in the UV regime and describes the dependence on $n$
(provided that $n$ is large enough).

Therefore a given operator $A$ contributes to $Q(n)-Q(n-1)$ a term
proportional to 
\begin{equation}
\int_{0}^{1}\left(\frac{\Gamma(\alpha_{A}+n)}{\Gamma(\alpha_{A})\Gamma(n+1)}\right)^{2}r^{2n-1+2h_{V}}dr=\frac{1}{2(n+h_{V})}\left(\frac{\Gamma(\alpha_{A}+n)}{\Gamma(\alpha_{A})\Gamma(n+1)}\right)^{2}\label{eq:ndepgammafun}
\end{equation}
where 
\begin{equation}
\alpha_{A}=h_{\mathcal{O}}+h_{V}-h_{A}
\end{equation}
and all factors independent of $n$ were omitted. To evaluate the
leading \foreignlanguage{british}{behaviour} for large $n$, one can
use Stirling's formula 
\begin{equation}
\Gamma(x+1)\approx\sqrt{2\pi x}x^{x}e^{-x}\left(1+O(1/x)\right)
\end{equation}
and the relation 
\begin{equation}
(n+a)^{(n+a)}\approx n^{n+a}e^{a}\left(1+O(1/n)\right)
\end{equation}
which results in 
\begin{equation}
Q(n)-Q(n-1)\propto\sum_{A}K_{A}n^{2\alpha_{A}-3}\left(1+O(1/n)\right)
\end{equation}
where $K_{A}$ are $n$-independent coefficients. For large $n$,
this gives the differential equation%
\footnote{We remark that the difference equation is also easy to solve and the
difference from the differential equation is numerically important
for small $n$. On the other hand, for the relatively high values
of the cutoff the difference between the two is insignificant. %
} 
\begin{equation}
\frac{dQ(n)}{dn}\propto\sum_{A}K_{A}n^{2\alpha_{A}-3}\left(1+O(1/n)\right)
\end{equation}
Solving the equation with the initial condition at $n=\infty$, the
result is 
\begin{equation}
Q(n)=Q(\infty)+\sum_{A}\tilde{K}_{A}n^{2\alpha_{A}-2}\left(1+O(1/n)\right)+O(\lambda^{2})\label{eq:Qrunning}
\end{equation}
The leading corrections in the descendant level are of order $1/n$
(as indicated), arising from omitting $h_{V}$ besides $n$, and also
from the approximations used in Stirling's formula. 

For the complete expectation value $E_{\mathcal{O}}(n)$ one must
also compute the running of the normalization factor $\mathcal{N}(n)$.
This is the quantity same as $Q(n)$ for the choice when $\mathcal{O}$
is the identity. Substituting the identity operator for $\mathcal{O}$
in (\ref{eqn:deltaq1}) the expression becomes zero due to the presence
of the projector $\tilde{P}_{n}$, and so $\mathcal{N}(n)$ does not
run to first order in $\lambda$: 
\begin{equation}
\frac{d\mathcal{N}(n)}{dn}=O(\lambda^{2})
\end{equation}
Supposing that the expectation value we compute is convergent in TCSA
(i.e. $2\alpha_{A}-2<0$ for all $A$), the end result for the matrix
element is: 
\begin{equation}
E_{\mathcal{O}}(n)=E_{\mathcal{O}}(\infty)+\sum_{A}{\tilde{K}}_{A}n^{2\alpha_{A}-2}\left(1+O(1/n)\right)+O(\lambda^{2})\label{eq:VEVRGrunning}
\end{equation}
The coefficients in eqn. (\ref{eq:VEVRGrunning}) can also be evaluated
and have the following form
\begin{equation}
\tilde{K}_{A}=\frac{\lambda(2\pi)^{-1+2\alpha_{A}}C_{OV}^{A}\left\langle A\right\rangle _{L}}{\Gamma(\alpha_{A})^{2}(2-2\alpha_{A})}L^{2-2\alpha_{A}}\label{eq:tildeK}
\end{equation}
where $C_{OV}^{A}$ is the appropriate conformal operator product
coefficient, and $\left\langle A\right\rangle _{L}$ is the exact
finite volume vacuum expectation value of the operator $A$. The latter
can be replaced (up to exponential corrections in $L$) by the infinite
volume vacuum expectation value, resulting in a power law for the
volume dependence of these coefficients.

The issue now is whether this calculation is self-consistent, i.e.
whether the $O(\lambda^{2})$ terms are smaller than the $O(\lambda)$
terms we evaluated. Since they are rather complicated to evaluate,
we rather use a ``proxy'' condition which is to require that the
first order correction itself is small when compared to $E_{\mathcal{O}}(\infty)$.
One can neglect the exponentially small difference between the infinite
and finite volume expectation values of $\mathcal{O}$: 
\begin{equation}
E_{\mathcal{O}}(\infty)-\mbox{\ensuremath{\left\langle \mathcal{O}\right\rangle }}_{\lambda}\sim O\left(e^{-mL}\right)
\end{equation}
and assuming that $A$ is a primary field (so it transforms homogeneously
under the exponential mapping) we get 
\begin{equation}
\left.\left\langle \mbox{min}\right|U_{+}^{(\infty)}(\lambda)A(w,\bar{w})U_{-}^{(\infty)}(\lambda)\left|\mbox{min}\right\rangle \right|_{w=\bar{w}=1}=\left(\frac{L}{2\pi}\right)^{2h_{A}}\left\langle A\right\rangle _{\lambda}+O\left(e^{-mL}\right)
\end{equation}
in terms of the infinity volume vacuum expectation value $\left\langle A\right\rangle _{\lambda}$.
Up to some dimensionless numerical coefficients, this leads to the
condition 
\begin{equation}
\lambda\left(\frac{2\pi}{L}\right)^{2h_{\mathcal{O}}}\left(\frac{2\pi}{L}\right)^{2h_{V}}2\left(\frac{L}{2\pi}\right)^{2}\left(\frac{L}{2\pi}\right)^{2h_{A}}\left\langle A\right\rangle _{\lambda}n^{2\alpha_{A}-2}\ll\mbox{\ensuremath{\left\langle \mathcal{O}\right\rangle }}_{\lambda}
\end{equation}
Simple dimensional analysis gives 
\begin{equation}
\lambda\propto m^{2-2\Delta_{V}}\quad,\quad\left\langle A\right\rangle _{\lambda}\propto m^{2-2\Delta_{A}}\quad,\quad\mbox{\ensuremath{\left\langle \mathcal{O}\right\rangle }}_{\lambda}\propto m^{2-2\Delta_{\mathcal{O}}}
\end{equation}
up to numerical coefficients. This leads to 
\begin{equation}
\left(\frac{2\pi n}{mL}\right)^{2\alpha_{A}-2}\ll1\label{eq:PTfirstordercondition}
\end{equation}
i.e. 
\begin{equation}
\frac{mL}{2\pi n}\ll1
\end{equation}
which is the same condition as (\ref{eq:CPTcondition}). This is indeed
consistent with what we have observed in all of our practical calculations
performed so far: the first order extrapolation breaks down in large
enough volume (the precise value of which depends on the model considered);
and the higher the value of the cut-off, the larger is the critical
value of the volume where the break-down occurs.

\subsection{\label{sub:Remarks-on-TCSARG}Remarks}

First of all note that to leading order all the calculation is unchanged
if we consider a general matrix element instead of a vacuum expectation
value, as long as the cut-off is chosen so that the states between
which the matrix element is taken are included in the truncated Hilbert
space. The only difference is that instead of the vacuum matrix element
of operator $A$ one obtains the appropriate excited state finite-volume
matrix element as coefficient. Therefore, while the expression for
the exponent of the cutoff dependence is universal, its coefficient
depends on the matrix element considered.

If the operator is such that its matrix elements are not convergent
with increasing cutoff, the operator must be renormalized by subtracting
the divergent parts. These are always proportional to operators with
lower scaling dimensions, as long as the perturbation is relevant
(i.e. the theory is renormalizable). This is true because the condition
for divergence is 
\begin{equation}
2\alpha_{A}-2>0
\end{equation}
which leads to 
\begin{equation}
2(h_{\mathcal{O}}+h_{V}-h_{A})-2>0
\end{equation}
and since in a \foreignlanguage{british}{renormalizable} theory the
perturbing operator is relevant ($h_{V}\leq1$), a divergent term
can only be obtained if for some $A$ 
\begin{equation}
h_{A}<h_{\mathcal{O}}
\end{equation}
In this case, a renormalized operator $\left[\mathcal{O}\right]$
must be defined by subtracting from $\mathcal{O}$ a counter term
proportional to $A$. This is consistent with the general theorems
of renormalization derived in the framework of the standard Feynman
perturbation expansion in renormalizable quantum field theory.

Finally, note that the cut-off dependence can also be obtained by
a simple scaling argument, by examining the operator product expansion:
\begin{equation}
\int d^{2}zV\mathcal{O}\sim\sum_{A}C_{A}A
\end{equation}
The energy dependence of the coefficient is given by simple dimensional
analysis as 
\begin{equation}
C_{A}\propto[\mbox{energy}]^{-2+2\alpha_{A}}\quad,\quad\alpha_{A}=h_{\mathcal{O}}+h_{V}-h_{A}
\end{equation}
The typical energy of (zero-spin) states at descendant level $n$
is 
\begin{equation}
\frac{4\pi n}{L}
\end{equation}
If $n$ is high, then this energy is much larger than the mass scale
$m$, so by usual rules of ultraviolet perturbation theory the dependence
on the mass scale can be neglected to leading order, which means that
the above energy is the only scale in the calculation. Therefore the
contribution from level $n$ states to leading order takes the form
\begin{equation}
C_{A}\propto\left(\frac{4\pi n}{L}\right)^{-2+2\alpha_{A}}
\end{equation}
which also gives the correct volume dependence as well.

In TCSA everything is measured in units of the mass scale $m$, so
the relevant dimensionless number which characterizes the magnitude
of the first order correction is 
\begin{equation}
c_{A}=C_{A}m^{2-2\alpha_{A}}\propto\left(\frac{4\pi n}{mL}\right)^{-2+2\alpha_{A}}
\end{equation}
in agreement with (\ref{eq:PTfirstordercondition}).

\subsection{Some particular models}

Here we consider four models that can be formulated as perturbed conformal
field theories. For the first two cases, we only present the exponents
as the relevant TCSA analysis was already discussed in previous publications.
For the other two cases, we also present the detailed numerical analysis
of the extrapolation procedure.

\subsubsection{Ising model in a magnetic field}

This is the model with the famous $E_{8}$ scattering theory obtained
by Zamolodchikov \cite{Zamolodchikov:1989fp}. The Hamiltonian is
given by 
\begin{equation}
H=H_{*}+\lambda\int dx\,\sigma
\end{equation}
where the role of the perturbing operator $V$ is played by the magnetization
(spin) field $\sigma$ which has $h_{\sigma}=\bar{h}_{\sigma}=1/16.$
The most relevant operators occurring in $\sigma\sigma$ are 
\begin{eqnarray}
A=1 & : & h_{A}=0\nonumber \\
A=\epsilon & : & h_{A}=1/2
\end{eqnarray}
and so the corresponding leading cut-off dependencies are 
\begin{equation}
n^{-7/4}\qquad,\qquad n^{-11/4}
\end{equation}
where the second exponential is of the same order as the $1/n$ corrections
to the first term. In fact, an extrapolation using the first exponent
gives an excellent result for the vacuum--two-particle matrix element,
as demonstrated in \cite{Konik:2007cb}.

\subsubsection{Sine-Gordon model}

The action of sine-Gordon model is 
\[
\mathcal{A}=\int dtdx\,\frac{1}{2}\left(\left(\partial_{t}\Phi\right)^{2}-\left(\partial_{x}\Phi\right)^{2}+\lambda\cos\beta\Phi\right)
\]
It can be described as a perturbation of a massless free boson ($c=1$
conformal field theory) with the Hamiltonian 
\begin{equation}
H=\int dx\,\frac{1}{2}\left(\left(\partial_{t}\Phi\right)^{2}+\left(\partial_{x}\Phi\right)^{2}\right)-\lambda\int dx\,:\cos\beta\Phi:
\end{equation}
(with the normal ordering according to the massless free boson modes),
and so in this case $V=-:\cos\beta\Phi:$. The conformal Hilbert space
is generated by vertex operators $W_{n,q}$ with $n,q\in\mathbb{Z}$,
where $q$ is called winding number and corresponds to the solitonic
topological charge. The $W_{n,q}$ have the conformal dimensions
\begin{align}
 & h_{n,q}=\frac{1}{2}\left(\frac{n}{R}+\frac{qR}{2}\right)^{2}\nonumber \\
 & \bar{h}_{n,q}=\frac{1}{2}\left(\frac{n}{R}-\frac{qR}{2}\right)^{2}\qquad R=\frac{\sqrt{4\pi}}{\beta}
\end{align}
The operators with $q=0$ have the simple form
\[
W_{n,0}=:e^{in\beta\Phi}:
\]
and therefore $V\equiv-\left(W_{1,0}+W_{-1,0}\right)$. To each $W_{n,m}$
there is an associated Fock module, whose basis is generated by the
action of the bosonic creation/annihilation operators. The TCSA was
first applied to the sine-Gordon model in \cite{Feverati:1998va}. 

At first sight our derivation of the cut-off dependence is not really
applicable here. As the spectrum of primary fields is not bounded
from above, with increasing cut-off $e_{\mathrm{cut}}$ new Fock modules
enter the Hilbert space. Therefore there is no simple relation between
the cut-off and the level of the highest descendants in the Hilbert
space. Nevertheless, it turns out that in the case discussed below
only two operators contribute in the operator product, and therefore
there is a one-to-one relation between $e_{\mathrm{cut}}$ and the
highest descendant level $n$ included in the two contributing Fock
modules, which is enough for our derivation to apply.

Let us investigate the operator

\begin{equation}
\mathcal{O}=:e^{i\beta\Phi}:
\end{equation}
We have 
\begin{equation}
h_{\mathcal{O}}=h_{V}=h=\frac{\beta^{2}}{8\pi}
\end{equation}
Using the operator product of exponential fields 
\begin{equation}
:e^{i\alpha\Phi(z,\bar{z})}::e^{i\beta\Phi(w,\bar{w})}:=|z-w|^{-\frac{\alpha\beta}{2\pi}}:e^{i\alpha\Phi(z,\bar{z})+}{}^{i\beta\Phi(w,\bar{w})}:
\end{equation}
the leading operators in the $\mathcal{O}V$ OPE are: 
\begin{eqnarray}
A=1 & : & h_{A}=0\\
A=\partial\Phi\bar{\partial}\Phi & : & h_{A}=1\nonumber \\
A=e^{2i\beta\Phi} & : & h_{A}=4h
\end{eqnarray}
which lead to the following cut-off dependence: 
\begin{equation}
n^{-2+4h}\qquad,\qquad n^{-4+4h}\qquad,\qquad n^{-2-4h}
\end{equation}
Note that for $8h>1$, the $1/n$ corrections to the first term have
a larger exponent than the third.

The extrapolation with only the leading term, when combined with a
numerical renormalization group, gives excellent results in the attractive
regime, as demonstrated in \cite{Palmai:2012kf}.

\subsubsection{Scaling Lee-Yang model \label{sub:Scaling-Lee-Yang-model-RGTCSA}}

\begin{figure}
\begin{centering}
\includegraphics{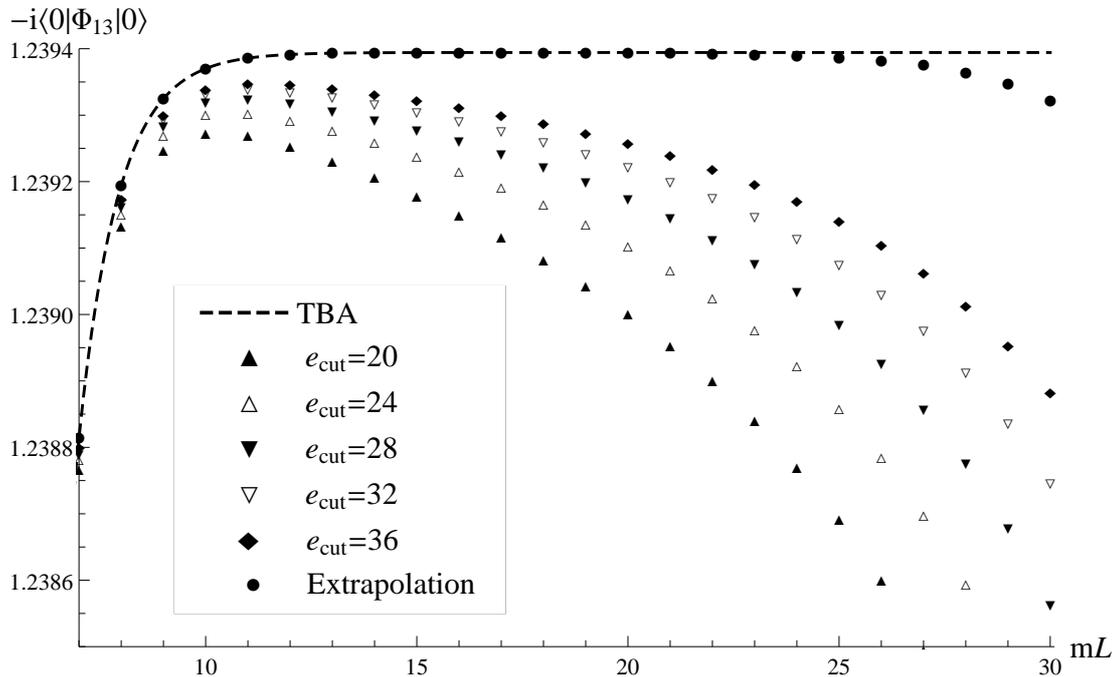} 
\par\end{centering}

\caption{\label{fig:Extrapolation-LY} Extrapolation of the dimensionless vacuum
expectation value $-im^{2/5}\left\langle \Phi_{1,3}\right\rangle _{L}$
in the scaling Lee-Yang model, compared to the exact value computed
from TBA.}
\end{figure}

The scaling Lee-Yang model is the perturbation of the $c=-22/5$ Lee-Yang
conformal field theory with the single relevant field $\Phi_{1,3}$
which has conformal weight $-1/5$, with the action
\begin{equation}
\mathcal{A}_{SLY}=\mathcal{A}_{LY}-\lambda\int dx\Phi_{1,3}\label{eq:lypcftham-2}
\end{equation}
where $\mathcal{A}_{LY}$ is the formal action of the Lee-Yang conformal
field theory. When $\lambda$ is a positive imaginary number, the
model has a single particle in its spectrum with mass $m$ that can
be related to the coupling constant as \cite{Zamolodchikov:1989cf}
\begin{equation}
\lambda=0.09704845636\dots\times i\, m^{12/5}\label{eq:lymassgap}
\end{equation}
and with a factorized scattering based on the two-particle S-matrix
\cite{Cardy1989} 
\begin{equation}
S(\theta)=\frac{\sinh\theta+i\sin\frac{2\pi}{3}}{\sinh\theta-i\sin\frac{2\pi}{3}}\label{eq:LYSmat}
\end{equation}
This is the model for which TCSA was originally developed by Yurov
and Zamolodchikov \cite{Yurov:1989yu}; for a detailed description
of our TCSA implementation we refer the reader to \cite{Pozsgay:2007kn}.
We choose $\Phi_{1,3}$ as the measured operator $\mathcal{O}$:

\begin{equation}
\mathcal{O}=\Phi_{1,3}
\end{equation}
which has the exact vacuum expectation value \cite{Fateev:1997yg}
\[
\left\langle \Phi_{1,3}\right\rangle =1.239394325\dots\times i\, m^{-2/5}
\]
in infinite volume. Given that $V=\Phi$, the relevant scaling dimensions
are 
\begin{equation}
h_{\mathcal{O}}=h_{V}=-\frac{1}{5}
\end{equation}
The operators occurring in the product $\mathcal{O}V=\Phi_{1,3}\Phi_{1,3}$
correspond to the family of the field $\Phi_{1,3}$ itself, and of
the identity, therefore the leading corrections result from 
\begin{eqnarray}
A=\Phi & : & h_{A}=-\frac{1}{5}\Rightarrow\alpha_{A}=-\frac{1}{5}\nonumber \\
A=\mathbf{1} & : & h_{A}=0\Rightarrow\alpha_{A}=-\frac{2}{5}
\end{eqnarray}
and so the cut-off dependence is given by the terms 
\begin{equation}
n^{-12/5}\qquad,\qquad n^{-14/5}
\end{equation}
When applying this cut-off dependence to extrapolate TCSA measurements
of the finite volume vacuum expectation value to infinite cut-off,
it turned out that TCSA data are so detailed and accurate that we
could also fit the subleading exponents 
\begin{equation}
n^{-17/5}\qquad,\qquad n^{-19/5}
\end{equation}
resulting from the $1/n$ corrections. These fits to the cutoff dependence
are very accurate, and the resulting extrapolated matrix elements
are shown in figure \ref{fig:Extrapolation-LY}. These results also
demonstrates that the RG improvement is really necessary to guarantee
a proper matching between TCSA and the Leclair-Mussardo series. In
this particular case the LM series is equivalent to the TBA derived
in \cite{Zamolodchikov:1989cf}). The construction is similar to the
case of $T_{2}$ discussed in \ref{sub:The-Leclair-Mussardo-series},
but is simpler as the factorized scattering theory of the model consists
of a single particle with the $S$ matrix ().

Note that, as predicted by the argument in Section \ref{sub:Remarks-on-TCSARG},
the extrapolation becomes inaccurate in large volume. One expects
that in this case it is necessary to evaluate the cutoff dependence
to second order in perturbation theory, which would require integrating
a four point function in the conformal field theory, and is out of
scope of the present work.

One can also check the coefficients $\tilde{K}_{A}$ of the leading
cutoff dependence exponents. We define real and dimensionless coefficients
by the relation 
\begin{equation}
\bar{K}_{A}=-i\tilde{K}_{A}m^{-2h_{\mathcal{O}}}
\end{equation}
which can be extracted from the cutoff dependence fits as functions
of the dimensionless volume $mL$. In view of the prediction (\ref{eq:tildeK}),
we then fit their volume dependence with a simple power 
\begin{equation}
\bar{K}_{A}(L)=k_{A}(mL)^{\beta_{A}}
\end{equation}
using $k_{A}$ and $\beta_{A}$ as fitting parameters. The fitted
values and the predictions are compared in Table \ref{tab:Cutoff-dependence-parameters-LeeYang};
they agree up to the numerical limitations. The numerical limitations
of TCSA come from neglecting $1/n$ terms in fitting the cut-off dependence;
the choices of the fitting windows, both in the cut-off and then in
the volume; from interference between the different powers of $n$
during fitting; and from numerical inaccuracies in matrix diagonalization.
The theoretical predictions also have systematic errors coming from
higher order corrections in perturbation theory, and from neglecting
the (exponential) volume dependence of the vacuum expectation values.
Despite all these limitations, it turns out that the agreement between
predicted and measured values of $\beta_{A}$ and $k_{A}$ is typically
within 10-20\% or better. 

\begin{table}
\begin{centering}
\begin{tabular}{|c|c|c|c|c|c|}
\hline 
Operator $A$ & $\beta_{A}$ (predicted) & $\beta_{A}$ (TCSA) & OPE $C_{OV}^{A}$ & $k_{A}$ (predicted) & $k_{A}$ (TCSA)\tabularnewline
\hline 
\hline 
$\Phi_{1,3}$ & $2.4$ & $2.41$ & $1.9113127\dots\times i$ & $-0.000215$ & $-0.000192$\tabularnewline
\hline 
$\mathbf{1}$ & $2.8$ & $2.83$ & $1$ & $0.0000914$ & $0.0000789$\tabularnewline
\hline 
\end{tabular}
\par\end{centering}

\caption{\label{tab:Cutoff-dependence-parameters-LeeYang} Cutoff dependence
parameters in the Lee-Yang model. The agreement between the predicted
and TCSA values is excellent for the exponents $\beta_{A}$. The coefficients
$k_{A}$ are more difficult to extract precisely from TCSA; also the
coefficient of the subleading exponent (corresponding to $\mathbf{1}$)
is harder to fit accurately due to its smaller value and the presence
of the leading $\Phi_{1,3}$ term.}

\end{table}

\subsubsection{$T_{2}$ model \label{sub:T2-model-RGTCSA}}

Let us first take the measured operator to be

\begin{equation}
\mathcal{O}=\Phi_{1,2}
\end{equation}
Since $V=\Phi_{1,3}$, according to (\ref{eq:t2operules}) the operators
occurring in the product $\mathcal{O}V$ are 
\begin{eqnarray}
A=\Phi_{1,2} & : & h_{A}=-\frac{2}{7}\Rightarrow\alpha_{A}=-\frac{3}{7}\nonumber \\
A=\Phi_{1,4}\equiv\Phi_{1,3} & : & h_{A}=-\frac{3}{7}\Rightarrow\alpha_{A}=-\frac{2}{7}
\end{eqnarray}
and so the cut-off dependence is given by 
\begin{equation}
n^{-18/7}\qquad,\qquad n^{-20/7}
\end{equation}
We again define real and dimensionless quantities with 
\begin{equation}
\bar{K}_{A}=i\tilde{K}_{A}m_{1}^{-2h_{\mathcal{O}}}
\end{equation}
and fit
\begin{equation}
\bar{K}_{A}(L)=k_{A}(m_{1}L)^{\beta_{A}}
\end{equation}
The measured values and the predictions for $k_{A}$ and $\beta_{A}$
are compared in Table \ref{tab:Cutoff-dependence-parameters-T2-12}.
The agreement of the extrapolated expectation value to theoretical
predictions is analyzed in section \ref{sec:Numerical-comparison}.
In the case of the $T_{2}$ model a further source of error appears,
namely the extraction of the $\bar{K}_{A}$ from the cut-off dependence
is itself plagued with fitting errors, and therefore the fit of their
$L$-dependence is even less accurate. As a result, the uncertainties
resulting from the fitting are also displayed in tables \ref{tab:Cutoff-dependence-parameters-T2-12},
\ref{tab:Cutoff-dependence-parameters-T2-13} and \ref{tab:Cutoff-dependence-parameters-T2-13new}.
(In contrast, the values obtained for the Lee-Yang model do not have
any fitting uncertainties affecting any of the digits displayed in
Table \ref{tab:Cutoff-dependence-parameters-LeeYang}).

\begin{figure}
\begin{centering}
\includegraphics{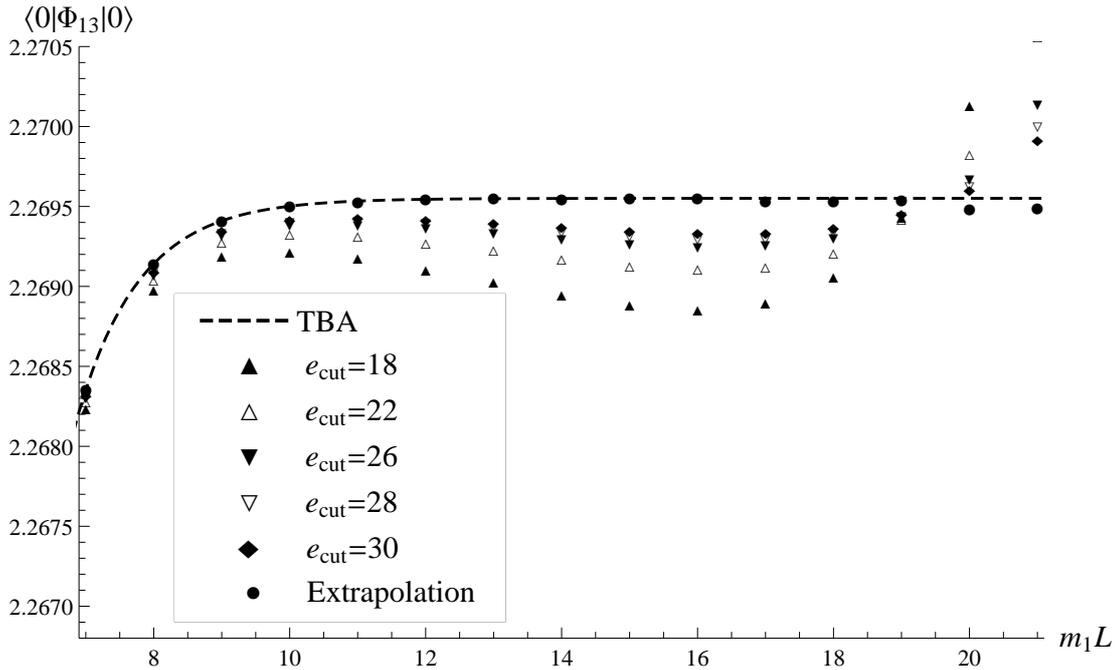} 
\par\end{centering}

\caption{\label{fig:Extrapolation-T2} Extrapolation of the dimensionless vacuum
expectation value $m^{6/7}\left\langle \Phi_{1,3}\right\rangle $
in the $T_{2}$ model, compared to the exact value computed from TBA.}
\end{figure}

\begin{table}
\begin{centering}
\begin{tabular}{|c|c|c|c|c|c|}
\hline 
Operator $A$ & $\beta_{A}$ (predicted) & $\beta_{A}$ (TCSA) & OPE $C_{OV}^{A}$ & $k_{A}$ (predicted) & $k_{A}$ (TCSA)\tabularnewline
\hline 
\hline 
$\Phi_{1,3}$ & $2.571$ & $2.62(6)$ & $-4.592000\dots\times i$ & $-0.000459$ & $-0.00042(5)$\tabularnewline
\hline 
$\Phi_{1,2}$ & $2.857$ & $2.9(1)$ & $-2.569126\dots$ & $0.000211$ & $0.00027(6)$\tabularnewline
\hline 
\end{tabular}
\par\end{centering}

\caption{\label{tab:Cutoff-dependence-parameters-T2-12} Cutoff dependence
parameters for the operator $\Phi_{1,2}$ in the $T_{2}$ model. The
agreement between the predicted and TCSA values is excellent for the
exponents $\beta_{A}$. The numbers in the parentheses are the uncertainties
of the last displayed digits resulting from the volume dependence
fits.}
\end{table}
For the case

\begin{equation}
\mathcal{O}=\Phi_{1,3}
\end{equation}
the operators occurring in the product $\mathcal{O}V$ are 
\begin{eqnarray}
A=1 & : & h_{A}=0\Rightarrow\alpha_{A}=-\frac{6}{7}\nonumber \\
A=\Phi_{1,3} & : & h_{A}=-\frac{3}{7}\Rightarrow\alpha_{A}=-\frac{3}{7}
\end{eqnarray}
and so the cut-off dependence is 
\begin{equation}
n^{-20/7}\qquad,\qquad n^{-26/7}
\end{equation}
The result of the extrapolation for the perturbing operator $\Phi_{1,3}$
is shown in figure \ref{fig:Extrapolation-T2}; again, the RG improvement
is really necessary to guarantee a proper agreement.

For the case of $\Phi_{1,3}$, one can define real and dimensionless
quantities with 
\begin{equation}
\bar{K}_{A}=\tilde{K}_{A}m^{6/7}
\end{equation}
and again fit
\begin{equation}
\bar{K}_{A}(L)=k_{A}(mL)^{\beta_{A}}
\end{equation}
For the cut-off dependence itself, there is a small surprise as shown
in Table \ref{tab:Cutoff-dependence-parameters-T2-13}. 
\begin{table}
\begin{centering}
\begin{tabular}{|c|c|c|c|c|c|}
\hline 
Operator $A$ & $\beta_{A}$ (predicted) & $\beta_{A}$ (TCSA) & OPE $C_{OV}^{A}$ & $k_{A}$ (predicted) & $k_{A}$ (TCSA)\tabularnewline
\hline 
\hline 
$\Phi_{1,3}$ & $2.857$ & $2.71(5)$ & $6.019309\dots$ & $-0.000483$ & $-0.00039(4)$\tabularnewline
\hline 
$\mathbf{1}$ & $3.714$ & $3.0(2)$ & $1$ & $-1.27\times10^{-6}$ & $0.00039(1)$\tabularnewline
\hline 
\end{tabular}
\par\end{centering}

\caption{\label{tab:Cutoff-dependence-parameters-T2-13} Cutoff dependence
parameters for the operator $\Phi_{1,3}$ in the $T_{2}$ model. The
agreement between the predicted and TCSA values is good for $\Phi_{1,3}$.
Note the disagreement between the identity prediction and the numbers
extracted from TCSA which is explained in the main text.}
\end{table}
The identity contribution clearly does not have the predicted exponent
for the volume dependence, and the predicted coefficient is also much
smaller than the measured one. However, note that the difference between
the two exponents is close to unity. This means that the $1/n$ correction
to the $n$-dependence of the $\Phi_{1,3}$ contribution has a very
a similar exponent to the $\mathbf{1}$ contributions: it scales with
the cutoff level as$n^{-27/7}$. Simply adding this exponent to the
other two does not work, as the cutoff fits become unstable due to
the presence of two close exponents. One can instead use the fact
that the identity contribution is predicted to be small, and fit the
cutoff dependence with the exponents
\begin{equation}
n^{-20/7}\qquad,\qquad n^{-27/7}
\end{equation}
The coefficient of the $1/n$ correction can be evaluated by expanding
the $n$-dependence in (\ref{eq:ndepgammafun}) to higher order, which
results in
\begin{equation}
\tilde{K}_{A}n^{-2+2\alpha_{A}}+\tilde{K}_{A}^{(1)}n^{-3+2\alpha_{A}}+O(n^{-4+2\alpha_{A}})
\end{equation}
where $\tilde{K}_{A}$ is given in (\ref{eq:tildeK}) and 
\begin{equation}
\tilde{K}_{A}^{(1)}=\frac{\lambda(2\pi)^{-1+2\alpha_{A}}C_{OV}^{A}\left\langle A\right\rangle _{L}(-h_{V}+\alpha_{A}(\alpha_{A}-1))}{\Gamma(\alpha_{A})^{2}(3-2\alpha_{A})}L^{2-2\alpha_{A}}
\end{equation}
Again we can substitute $\left\langle A\right\rangle _{L}$ with its
infinite volume value up to terms that decay exponentially in $L$.
The comparison of this prediction to the data is shown in Table \ref{tab:Cutoff-dependence-parameters-T2-13new};
it is clear from the data that the cut-off dependence data are now
explained well.

\begin{table}
\begin{centering}
\begin{tabular}{|c|c|c|c|c|c|}
\hline 
Operator $A$ & $\beta_{A}$ (predicted) & $\beta_{A}$ (TCSA) & OPE $C_{OV}^{A}$ & $k_{A}$ (predicted) & $k_{A}$ (TCSA)\tabularnewline
\hline 
\hline 
$\Phi_{1,3}$ & $2.857$ & $2.71(5)$ & $6.019309\dots$ & $-0.000483$ & $-0.00039(4)$\tabularnewline
\hline 
$\Phi_{1,3}$(subleading) & $2.857$ & $3.0(2)$ & $6.019309\dots$ & $0.000372$ & $0.00039(1)$\tabularnewline
\hline 
\end{tabular}
\par\end{centering}

\caption{\label{tab:Cutoff-dependence-parameters-T2-13new} Cutoff dependence
parameters for the operator $\Phi_{1,3}$ in the $T_{2}$ model, fitting
only the $\Phi_{1,3}$ contribution, including its $1/n$ correction
(subleading term). }
\end{table}

\section{Numerical comparison \label{sec:Numerical-comparison}}

Here we aim to compare the Leclair-Mussardo series to the RG-extrapolated
TCSA in the $T_{2}$ model, and for the operators $\Phi_{1,3}$ and
$\Phi_{1,2}$. 

The $\Phi_{1,3}$ operator in the $T_{2}$ model is related to the
trace of the stress-energy tensor (\ref{eq:Theta_Phi13}), so its
finite volume expectation value can be evaluated from TBA and it does
not give us an independent validation for the LM series. However,
the error in the numerical solution of the TBA equation (\ref{tab:TBA_TCSA_Phi13})
can be made very small (in our case it was of the order $10^{-14}$)
and therefore this can be used to check our procedures for evaluation
of the LM series; we found excellent agreement. Another application
is to estimate the accuracy of the RG-extrapolated TCSA. Table \ref{tab:TBA_TCSA_Phi13}
shows the difference of the expectation value for the $\Phi_{1,3}$
operator between the TBA prediction and RG-extrapolated TCSA evaluation
in units of $m_{1}$ for several values of the dimensionless volume
parameter, $l=m_{1}L$. The results also shows the efficiency of the
RG-extrapolated TCSA by comparing the RG-extrapolated TCSA value to
those evaluated at the highest value of the cutoff. As exemplified
by the data at $l=15$, the extrapolation results of an improvement
of accuracy of almost two orders of magnitude for volumes $l\gtrsim10$.

\begin{table}[h]
\begin{centering}
\begin{tabular}{|c||c|c|c|c|c|c|c|}
\hline 
$l$  & $0.2$  & $0.5$  & $1$  & $2$  & $4$  & $6$  & $15$\tabularnewline
\hline 
$\delta f_{13}^{RG}$  & $8.6\cdot10^{-11}$  & $1.2\cdot10^{-8}$  & $5.3\cdot10^{-8}$  & $1.7\cdot10^{-7}$  & $5.5\cdot10^{-7}$  & $1.5\cdot10^{-6}$  & $3.6\cdot10^{-6}$ \tabularnewline
\hline 
$\delta f_{13}^{e_{cut}=30}$  & $8.6\cdot10^{-11}$  & $4.1\cdot10^{-9}$  & $1.0\cdot10^{-7}$  & $9.0\cdot10^{-7}$  & $6.5\cdot10^{-6}$  & $1.9\cdot10^{-5}$  & $1.9\cdot10^{-4}$ \tabularnewline
\hline 
\end{tabular}
\par\end{centering}

\caption{\label{tab:TBA_TCSA_Phi13}The difference $\delta f_{13}=m_{1}^{6/7}\left(\left\langle \Phi_{1,3}\right\rangle _{\mathrm{TBA}}-\left\langle \Phi_{1,3}\right\rangle _{\mathrm{TCSA}}\right)$
between the RG-extrapolated TCSA evaluation and the TBA prediction
for the VEV of $\Phi_{1,3}$. The second row shows the difference
between the raw TCSA value at the cut-off value $e_{cut}=30$ and
the TBA. }
\end{table}

Now we turn to the evaluation of the LM-series for the $\Phi_{1,2}$
operator. At infinite volume only the infinite volume VEV (\ref{eq:t2vevs})
contributes to the LM-series, while for small volumes almost all terms
are relevant. Their contributions are proportional to $\sim e^{-\mu L}$,
where $\mu$ is the sum of the masses of the particles entering a
given term. For every contribution there's a characteristic volume
where it becomes relevant. We calculated the first seven finite volume
correction terms: $1$, $2$, $11$, $12$, $111$, $22$ and $112$,
where the number of $1$ and $2$ means the number of particle with
mass $m_{1}$ and $m_{2}$ participating in the given contributions.
Table \ref{tab:Diff_TCSA_LM_Phi12_Overall} shows the value of the
RG-extrapolated TCSA evaluation and the LM-series with the mentioned
contributions for $\Phi_{1,2}$, while Table \ref{tab:Diff_TCSA_LM_Separate}
shows the difference of the LM-series from the RG-extrapolated TCSA
data while adding more and more contributions.

For $l>1$, there is a steady improvement as more and more terms are
added, and it is clear from the table that contributions from higher
states are switched on at progressively lower volumes. Comparing the
columns labeled $+111$ and $+12$, it can be seen that adding the
$111$ contribution makes the agreement of the LM series with TCSA
worse; the reason is that the term $22$ is of the same order of magnitude
as the contribution from $111$, so consistency requires adding them
to the series together. Indeed, the deviation reported in column $+22$
is smaller than the one in column$+12$.

\begin{table}[h]
\begin{centering}
$\begin{array}{|r||r|r|c|c|c|c|}
\hline l & 0.5 & 1. & 1.5 & 2. & 2.5 & 3.\\
\hline \mbox{TCSA} & 1.0776 & 1.5704 & 1.8957 & 2.0949 & 2.2060 & 2.2640\\
\hline \mbox{LM} & 1.3853 & 1.6215 & 1.9036 & 2.0960 & 2.2061 & 2.2640\\
\hline\hline l & 3.5 & 4. & 4.5 & 5. & 7. & 10.\\
\hline \mbox{TCSA} & 2.2936 & 2.3086 & 2.3163 & 2.3203 & 2.3246 & 2.3251\\
\hline \mbox{LM} & 2.2936 & 2.3086 & 2.3163 & 2.3203 & 2.3247 & 2.3251
\\\hline \end{array}$ 
\par\end{centering}

\caption{\label{tab:Diff_TCSA_LM_Phi12_Overall}The values for $im_{1}^{4/7}\left\langle \Phi_{1,2}\right\rangle $
from the RG-extrapolated TCSA evaluation and the Leclair-Mussardo
series. }
\end{table}

\begin{table}[h]
\[
\begin{array}{|r||r|r|r|r|r|r|r|r|}
\hline l & 0 & +1 & +2 & +11 & +12 & +111 & +22 & +112\\
\hline\hline 0.5 & -1.2475 & -0.7261 & 0.8304 & 0.7290 & 0.1691 & 0.1879 & -0.4571 & -0.3077\\
\hline 1 & -0.7547 & -0.4576 & 0.2609 & 0.2244 & 6\cdot10^{-2} & 7\cdot10^{-2} & -8\cdot10^{-2} & -5\cdot10^{-2}\\
\hline 1.5 & -0.4295 & -0.2530 & 8\cdot10^{-2} & 7\cdot10^{-2} & 2\cdot10^{-2} & 2\cdot10^{-2} & -1\cdot10^{-2} & -8\cdot10^{-3}\\
\hline 2 & -0.2302 & -0.1259 & 2\cdot10^{-2} & 2\cdot10^{-2} & 4\cdot10^{-3} & 5\cdot10^{-3} & -2\cdot10^{-3} & -1\cdot10^{-3}\\
\hline 2.5 & -0.1192 & -6\cdot10^{-2} & 6\cdot10^{-3} & 4\cdot10^{-3} & 9\cdot10^{-4} & 1\cdot10^{-3} & -3\cdot10^{-4} & -1\cdot10^{-4}\\
\hline 3 & -6\cdot10^{-2} & -3\cdot10^{-2} & 2\cdot10^{-3} & 1\cdot10^{-3} & 2\cdot10^{-4} & 2\cdot10^{-4} & -4\cdot10^{-5} & -1\cdot10^{-5}\\
\hline 3.5 & -3\cdot10^{-2} & -1\cdot10^{-2} & 5\cdot10^{-4} & 3\cdot10^{-4} & 3\cdot10^{-5} & 4\cdot10^{-5} & -6\cdot10^{-6} & -2\cdot10^{-6}\\
\hline 4 & -2\cdot10^{-2} & -5\cdot10^{-3} & 1\cdot10^{-4} & 6\cdot10^{-5} & 6\cdot10^{-6} & 6\cdot10^{-6} & -2\cdot10^{-6} & -1\cdot10^{-6}\\
\hline 4.5 & -9\cdot10^{-3} & -2\cdot10^{-3} & 4\cdot10^{-5} & 1\cdot10^{-5} & 2\cdot10^{-8} & 1\cdot10^{-7} & -1\cdot10^{-6} & -1\cdot10^{-6}\\
\hline 5 & -5\cdot10^{-3} & -9\cdot10^{-4} & 1\cdot10^{-5} & 3\cdot10^{-6} & -9\cdot10^{-7} & -9\cdot10^{-7} & -1\cdot10^{-6} & -1\cdot10^{-6}\\
\hline 7 & -5\cdot10^{-4} & -3\cdot10^{-5} & -1\cdot10^{-6} & -1\cdot10^{-6} & -1\cdot10^{-6} & -1\cdot10^{-6} & -1\cdot10^{-6} & -1\cdot10^{-6}\\
\hline 10 & -2\cdot10^{-5} & -2\cdot10^{-6} & -2\cdot10^{-6} & -2\cdot10^{-6} & -2\cdot10^{-6} & -2\cdot10^{-6} & -2\cdot10^{-6} & -2\cdot10^{-6}\\
\hline 15 & 8\cdot10^{-7} & 9\cdot10^{-7} & 9\cdot10^{-7} & 9\cdot10^{-7} & 9\cdot10^{-7} & 9\cdot10^{-7} & 9\cdot10^{-7} & 9\cdot10^{-7}
\\\hline \end{array}
\]

\caption{\label{tab:Diff_TCSA_LM_Separate}The difference $\delta f_{12}=im_{1}^{4/7}\left(\left\langle \Phi_{1,2}\right\rangle _{\mathrm{LM}}-\left\langle \Phi_{1,2}\right\rangle _{\mathrm{TCSA}}\right)$
between the RG-extrapolated TCSA evaluation and the Leclair-Mussardo
series, depending on the multi-particle contributions included in
the latter.}
\end{table}

For $l\lesssim1$, the series does not converge very well; indeed
for $l=0.5$, there is no sign of any convergence. However, for such
small values of the volume higher terms of the LM series would still
be significant. Indeed, the leading corrections following the $112$
term are the contributions $1111$ and $122$, which can be estimated
to be of order 
\begin{equation}
e^{-4m_{1}L}\quad\mbox{and}\quad e^{-(m_{1}+2m_{2})L}
\end{equation}
which at $l=0.5$ give approximately $14\%$ and $12\%$, respectively.
This agrees well with the magnitude of the deviation at $l=0.5$,
as can also be seen from table \ref{tab:Diff_TCSA_LM_Phi12_Overall}.

For volumes $l<3.5$ the estimated TCSA error is smaller than $10^{-6}$,
and so the deviation between TCSA and the LM series is dominated by
the higher corrections to the LM series. However, the calculation
of higher contributions becomes progressively slower as the number
of particles to include increases ($122$ can be evaluated using a
$10$-particle form factor, cf. the remark at the end of Appendix
\ref{sec:Scattering-theory-of-T2}). Similarly to the case of $111$
and $22$, contributions $1111$ and $122$ are roughly of the same
order and so they must be added to the series together. Treating the
$10$-particle form factor numerically proved to be rather difficult,
so we have not evaluated the contribution $122$.

\begin{figure}
\begin{centering}
\includegraphics{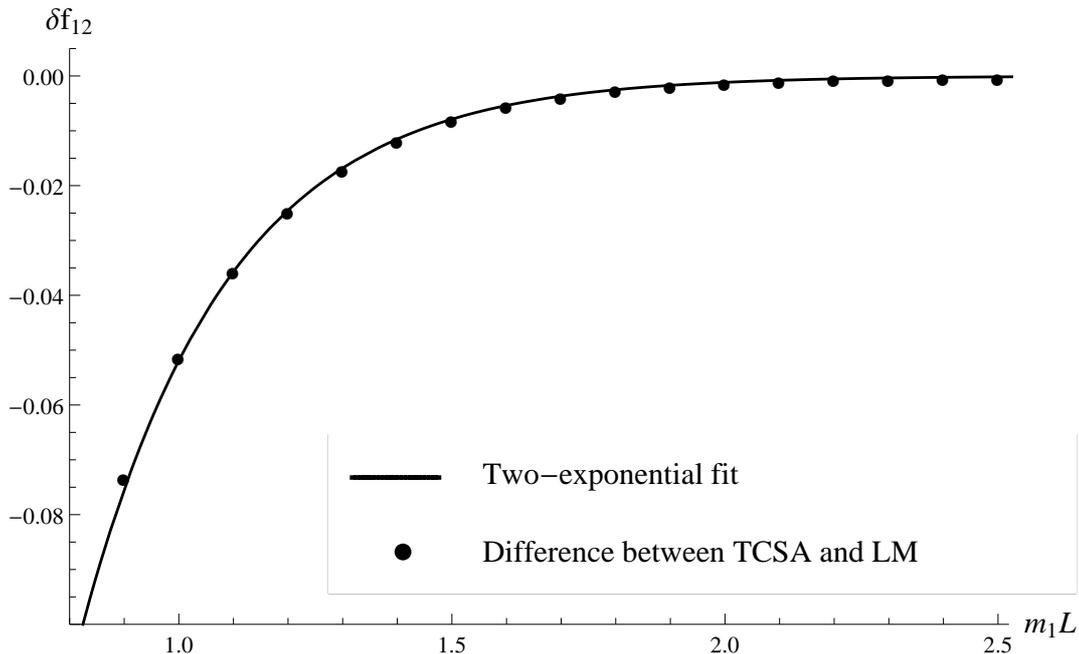} 
\par\end{centering}

\caption{\label{fig:Fitting-the-residual} Fitting the residual difference
$\delta f_{12}$ between the LM series evaluated up to $112$ and
the RG extrapolated TCSA result }
\end{figure}

However, there is a way to verify the consistency of the above considerations
further. We fitted the deviation in the last column of Table \ref{tab:Diff_TCSA_LM_Separate}
the difference from the TCSA by the Ansatz 
\begin{equation}
a\, e^{-4m_{1}L}+b\, e^{-(m_{1}+2m_{2})L}
\end{equation}
dictated by the particle content of the states $1111$ and $122$.
As shown in figure \ref{fig:Fitting-the-residual}, the fit was very
successful. In addition, the parameters $a$ and $b$ turned out to
be of the same order of magnitude, as expected from the above considerations.

\begin{figure}[h]
\begin{centering}
\includegraphics{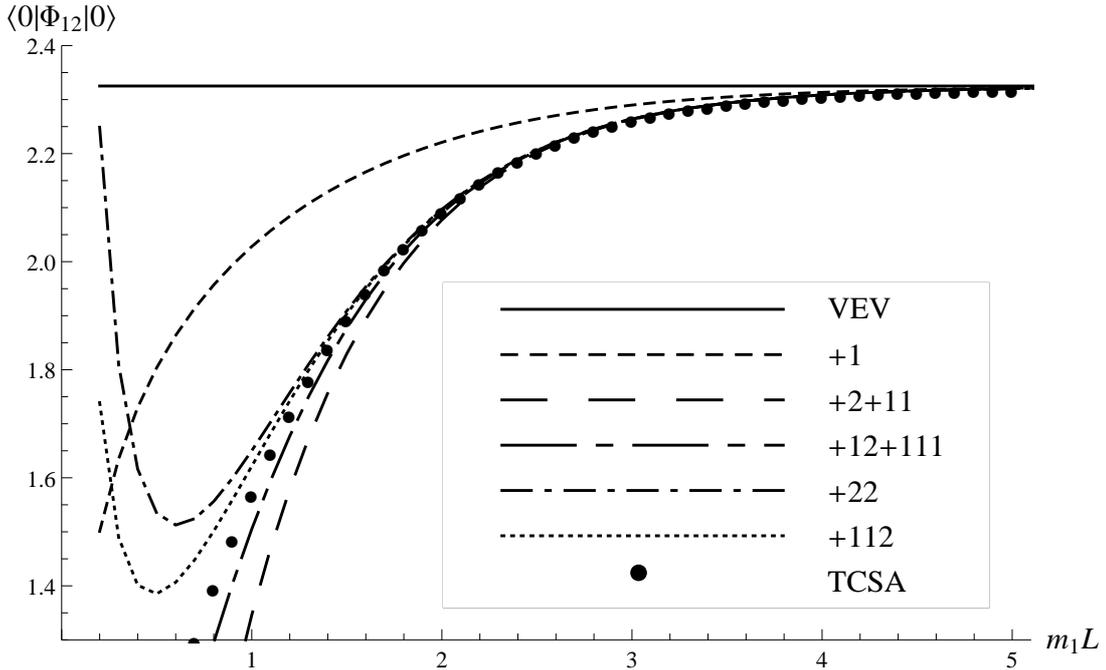} 
\par\end{centering}

\caption{\label{fig:LM_TCSA_comparison}Leclair-Mussardo series against RG-extrapolated
TCSA evaluation for the VEV of $\Phi_{1,2}$ by taking account more
and more contributions, in units of $m_{1}$}
\end{figure}

For larger volumes the agreement cannot be improved by including other
contributions, since it is dominated by the residual truncation error
of the RG-extrapolated TCSA. We remark that a similar evaluation of
the LM series for the case of $\Phi_{1,3}$ operator the agreement
of the LM series and TBA calculations gave an agreement of precision
$10^{-13}$ with the TBA for volumes $l>7$. Figure \ref{fig:LM_TCSA_comparison}
illustrates the improvement of the LM-series by adding more terms,
as compared to the RG-extrapolated TCSA data.

\section{A remark on finite volume form factors and $\mu$-terms \label{sec:A-remark-on-finite-volume-form-factors-and-mu-terms}}

In this section we digress to report an issue we found when we tested
the accuracy of TCSA by comparing numerical results for the finite
volume matrix elements to predictions of the finite volume form factor
formalism, using the methods developed in \cite{Pozsgay:2007kn,Pozsgay:2007gx}.

In \cite{Pozsgay:2007kn} Pozsgay and Takács derived a formula for
finite volume form factors valid up to exponential corrections: 
\begin{eqnarray}
 & _{j_{1},\dots,j_{n}}\left\langle \left\{ J_{1},\dots,J_{n}\right\} \right|\mathcal{O}\left(0,0\right)\left|\left\{ I_{1},\dots,I_{m}\right\} \right\rangle _{i_{1},\dots,i_{m};L}=\nonumber \\
 & \pm\frac{F_{j_{n},\dots,j_{1},i_{1},\dots,i_{m}}^{\mathcal{O}}\left(\theta_{n}'+i\pi,\dots,\theta_{1}'+i\pi,\theta_{1},\dots,\theta_{m}\right)}{\sqrt{\rho_{j_{1},\dots,j_{n}}\left(\theta_{1}',\dots,\theta_{n}'\right)\rho_{i_{1},\dots,i_{m}}\left(\theta_{1},\dots,\theta_{m}\right)}}\nonumber \\
 & \times\Phi_{j_{1},\dots,j_{n}}\left(\theta_{1}',\dots,\theta_{n}'\right)^{*}\Phi_{i_{1},\dots,i_{m}}\left(\theta_{1},\dots,\theta_{m}\right) & +O\left(e^{-\mu L}\right)\label{eq:Finite_volume_FF}
\end{eqnarray}
where $I$ and $J$ are the quantum numbers, $i$ and $j$ are the
particle types in the given states, $\rho$ is the density of sates
and $F^{\mathcal{O}}$ is the infinite volume form factor, the phase
factors $\Phi$ have the form 
\begin{equation}
\Phi_{i_{1},\dots,i_{m}}\left(\theta_{1},\dots,\theta_{m}\right)=\sqrt{\prod_{{k,l=1\atop k<l}}^{n}S_{i_{k}i_{l}}(\theta_{k}-\theta_{l})}
\end{equation}
and the $\pm$ sign corresponds to the ambiguity in choosing the branch
of the square root functions. We note that the same ambiguity is present
in TCSA due to the fact that the choice of eigenvectors is not unique.
Formula (\ref{eq:Finite_volume_FF}) is valid if there are no disconnected
terms in the matrix element; for the treatment of disconnected terms
cf. \cite{Pozsgay:2007gx} .

\begin{figure}
\begin{centering}
\includegraphics{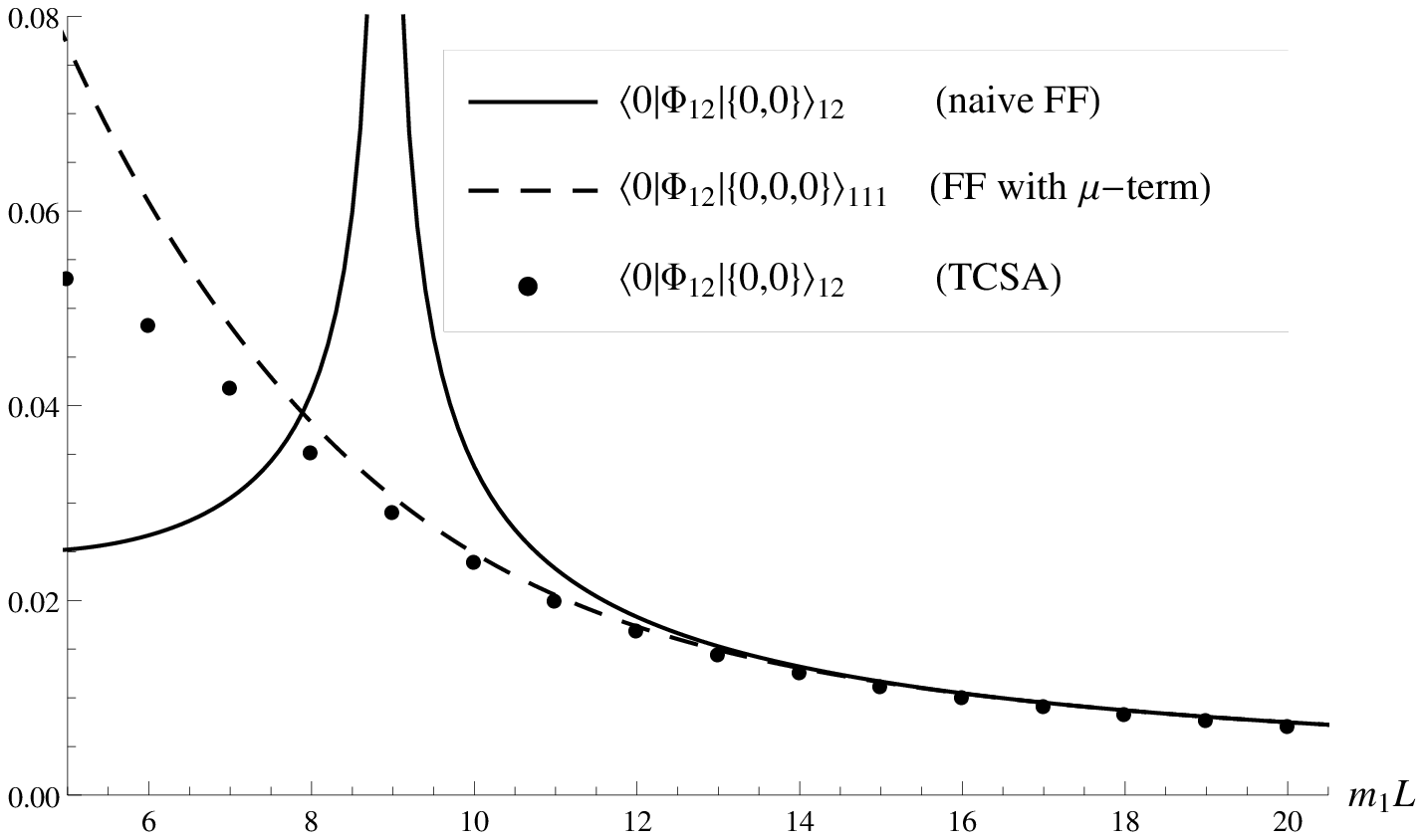}~\\
 \includegraphics{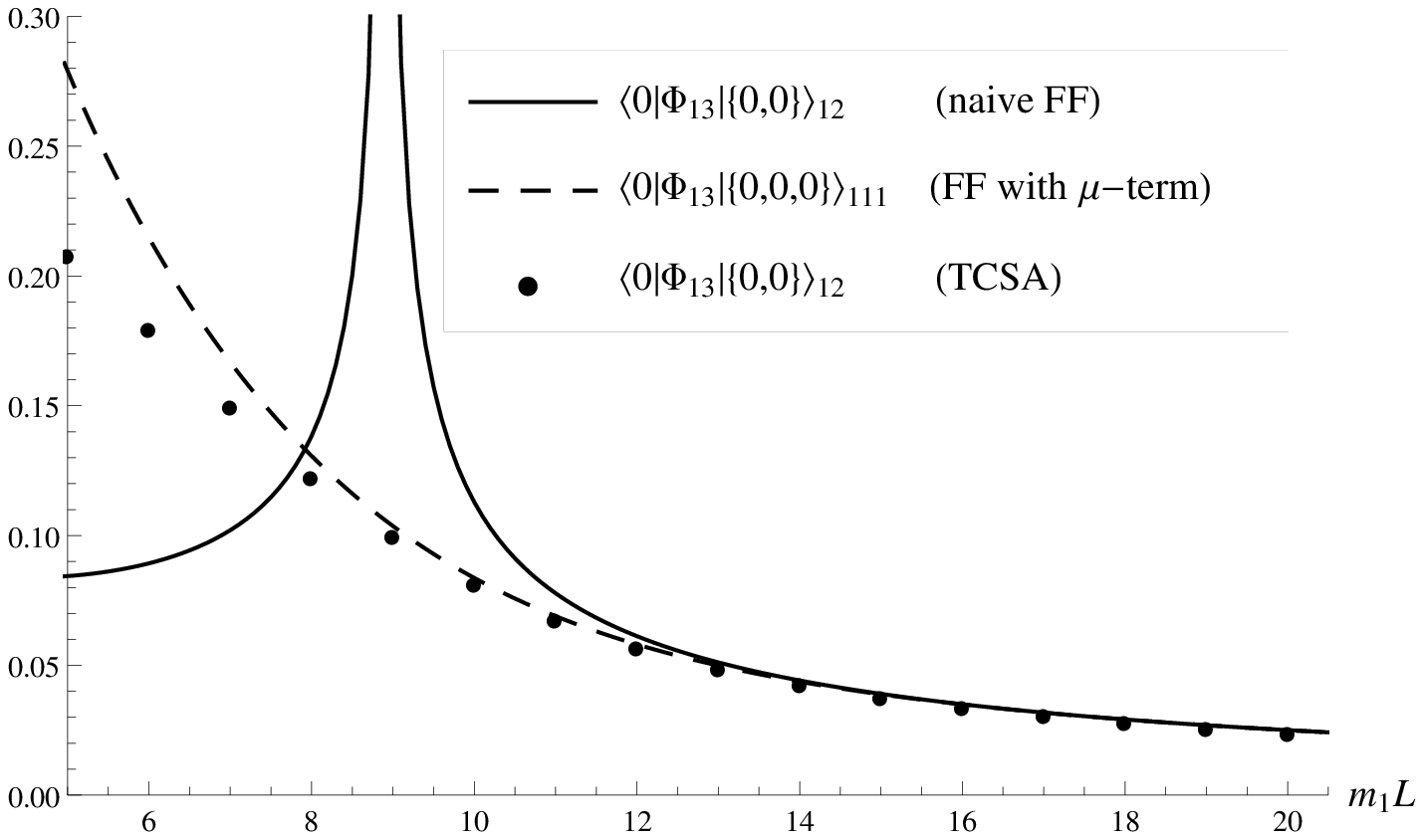} 
\par\end{centering}

\caption{\label{fig:muterm_vac_12}Vacuum-$B_{1}B_{2}$ matrix elements for
$\Phi_{1,2}$ and $\Phi_{1,3}$ operators in units of $m_{1}$, where
the solid line is the naive finite volume form factors result, the
dashed line is the correction with the $\mu$-terms and the dots represent
the TCSA data. All data in the plot show absolute values of the matrix
elements to get rid of the phase ambiguities related to choice of
eigenvectors in TCSA.}
\end{figure}

The density of states in (\ref{eq:Finite_volume_FF}) can be obtained
from the finite volume quantization conditions (valid up to exponential
corrections) 
\begin{eqnarray}
Q_{i_{1}\dots i_{n}}^{(k)}(\theta_{1},\dots,\theta_{n}) & = & m_{i_{k}}L\sinh\theta_{k}+\sum_{j\neq k}\delta_{i_{k}i_{j}}\left(\theta_{k}-\theta_{j}\right)=2\pi I_{k}\qquad\label{eq:breatherby}\\
 &  & I_{k}\in\mathbb{Z}\qquad,\qquad k=1,\dots,n\nonumber 
\end{eqnarray}
as 
\begin{equation}
\rho_{i_{1}\dots i_{n}}(\theta_{1},\dots,\theta_{n})=\det\mathcal{J}^{(n)}\qquad,\qquad\mathcal{J}_{kl}^{(n)}=\frac{\partial Q_{i_{1}\dots i_{n}}^{(k)}(\theta_{1},\dots,\theta_{n})}{\partial\theta_{l}}\quad,\quad k,l=1,\dots,n\label{eq:byjacobian}
\end{equation}
where the phase-shifts are defined as 
\begin{equation}
\delta_{i_{1}i_{2}}\left(\theta_{1}-\theta_{2}\right)=-i\log S_{i_{1}i_{2}}\left(\theta_{1}-\theta_{2}\right)
\end{equation}
The densities $\rho$ depend on the volume, and are positive for large
volumes since their leading behavior is given by 
\begin{equation}
\rho_{i_{1},\dots,i_{n}}\left(\theta_{1},\dots,\theta_{n}\right)=\prod_{k=1}^{n}m_{i_{k}}L\cosh\theta_{k}+O(L^{n-1})
\end{equation}
where the subleading corrections depend on the scattering phase shifts
\cite{Pozsgay:2007kn}. However, nothing prevents the $\rho$ functions
from becoming zero or negative for smaller values of the volume, which
would therefore produce a singularity in the finite volume form factor
predicted by (\ref{eq:Finite_volume_FF}). Such singularities have
not been encountered previously; however, in the $T_{2}$ model we
found this behaviour for the state $B_{1}B_{2}$, as shown in figure
\ref{fig:muterm_vac_12}.

On the other hand, the matrix elements calculated from TCSA prove
to be regular (no singularities)%
\footnote{For a non-unitary model this is not necessarily the case, as the eigenvectors
can have zero norm for some values of the volume. These norms appear
in the denominator when evaluating the matrix elements in TCSA \cite{Pozsgay:2007kn}.
This does happen in the case of the boundary Lee-Yang model when the
ground state level and the first excited level cross \cite{Dorey:2000eh}.%
}. Therefore we should take a closer look at the finite volume form
factor formula (\ref{eq:Finite_volume_FF}). Since it is correct only
up to exponential corrections, the natural idea is that exponential
corrections could resolve the singularities.

In \cite{Pozsgay:2008bf} Pozsgay introduced a method to describe
a particular class of exponential corrections, the so-called $\mu$-terms
which originate from the composite structure of the particles under
the bootstrap principle. Using the bound-state quantization first
introduced in \cite{Kausch:1996vq}, these can be described by a continuation
of the quantization relations (\ref{eq:breatherby}) and the form
factor formula (\ref{eq:Finite_volume_FF}) to complex rapidities.
For more details we refer the reader to the original article \cite{Pozsgay:2008bf},
and to \cite{Feher:2011aa,Takacs:2011nb} where this method was used
with great success.

To demonstrate how the exponential corrections dissolve the singularities,
let us examine the $\left|\left\{ 0,0\right\} \right\rangle _{12}$
state, which contains a $B_{1}$ and a $B_{2}$ particle, both with
quantum number $0$. The $B_{2}$ particle can be treated as the bound
state of two $B_{1}$ particles which have complex rapidity. We can
therefore treat the state as a three-$B_{1}$ state$\left|\left\{ 0,0,0\right\} \right\rangle _{111}$
and look for a solution of (\ref{eq:breatherby}) with rapidities
$\theta_{1}=0$, $\theta_{2}=iu$ and $\theta_{3}=-iu$. The single
unknown $u$ can be obtained by solving 
\begin{equation}
im_{1}L\sin u+\delta_{11}(iu)+\delta_{11}(2iu)=0
\end{equation}
and the analytic continuation of (\ref{eq:Finite_volume_FF}) (including
the phase factors) gives the finite volume form factor including the
$\mu$-term contribution. Figure \ref{fig:muterm_vac_12} shows the
finite volume form factor with the naive evaluation and the correction
by the $\mu$-terms against the TCSA data for the vacuum-$B_{1}B_{2}$
matrix elements. It can be seen clearly that the exponential corrections
do resolve the singularities, and the agreement between the TCSA data
and form factor expression is very good down to small volumes where
other exponential corrections become relevant.

\begin{figure}
\begin{centering}
\includegraphics{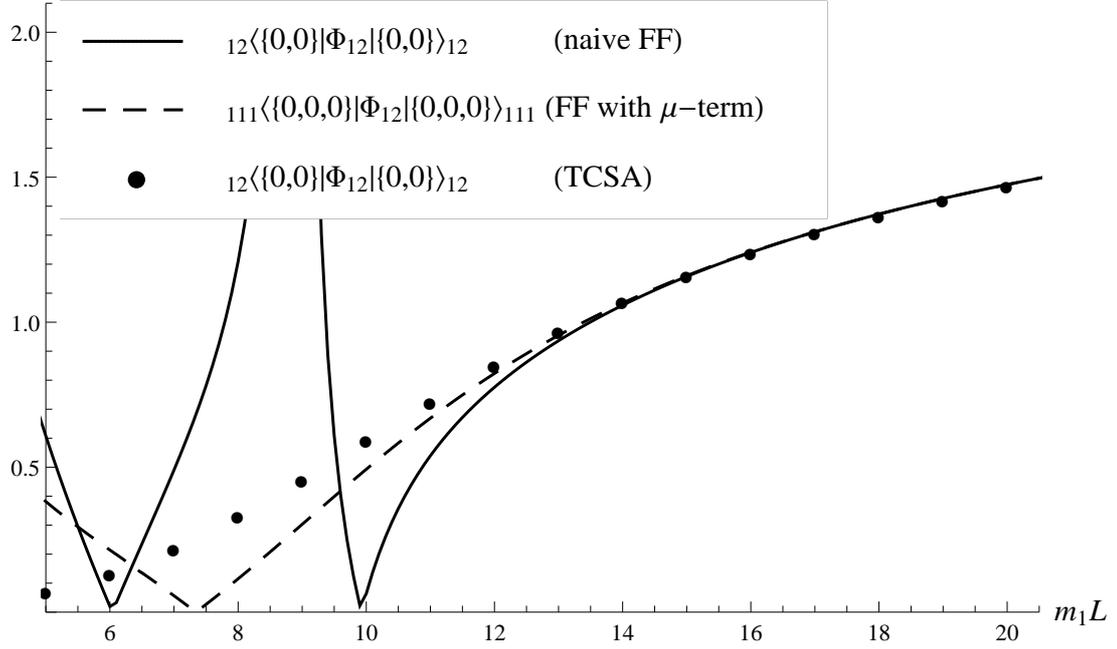}~\\
 \includegraphics{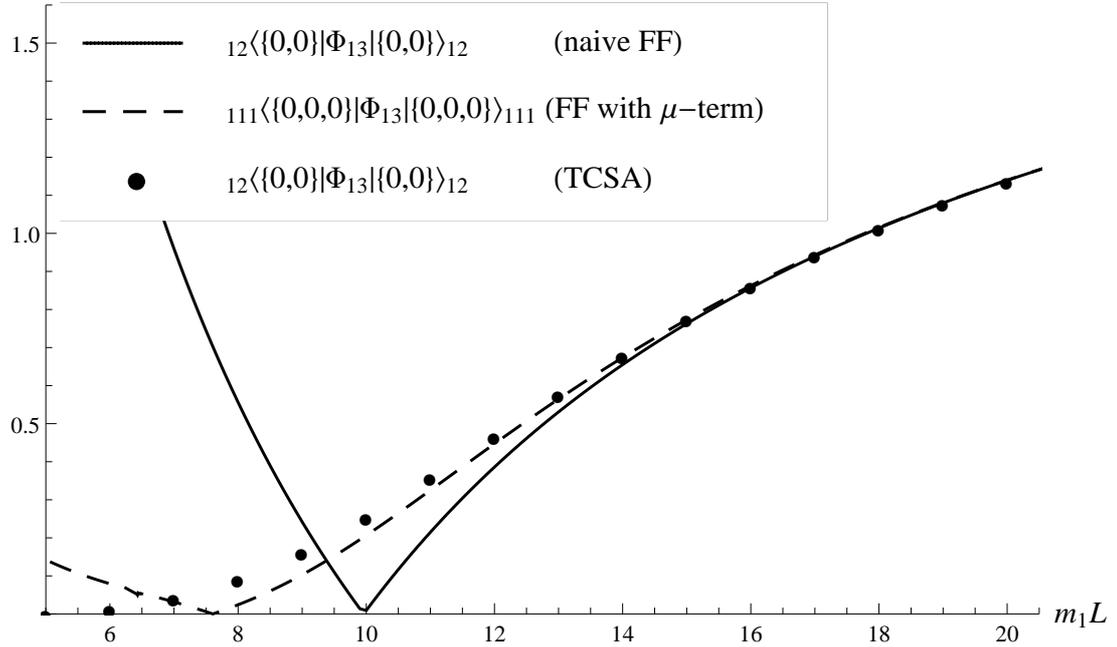} 
\par\end{centering}

\caption{\label{fig:muterm_12_12}$B_{1}B_{2}$-$B_{1}B_{2}$ matrix elements
for $\Phi_{1,2}$ and $\Phi_{1,3}$ operators in units of $m_{1}$,
where the solid line is the naive finite volume form factors result,
the dashed line is the correction with the $\mu$-terms and the dots
represent the TCSA data. The breaks in the lines are due to plotting
the absolute values of the predicted matrix elements (the true value
would become negative and the line cross below the real axis).}
\end{figure}

Figure \ref{fig:muterm_12_12} shows the results for the $B_{1}B_{2}$-$B_{1}B_{2}$
diagonal matrix elements%
\footnote{Note that these matrix elements contain disconnected contributions,
and therefore must be computed by a formula different from (\ref{eq:Finite_volume_FF});
for details cf. \cite{Pozsgay:2007gx,Pozsgay:2008bf,Takacs:2011nb}. %
}. The diagonal matrix element of $\Phi_{1,3}$ operator contains no
singularity, since it is related to the trace of the energy momentum
tensor, and its diagonal matrix element can be expressed using the
energy and momentum eigenvalues of the state as computed in the approximation
given by (\ref{eq:Finite_volume_FF}), which are finite. However even
in this case including the $\mu$-terms improves the agreement with
TCSA. For the operator $\Phi_{1,2}$ the conclusion is the same as
in the non-diagonal case: the $\mu$-terms again lead to a resolution
of the singularity.

\begin{figure}
\begin{centering}
\includegraphics{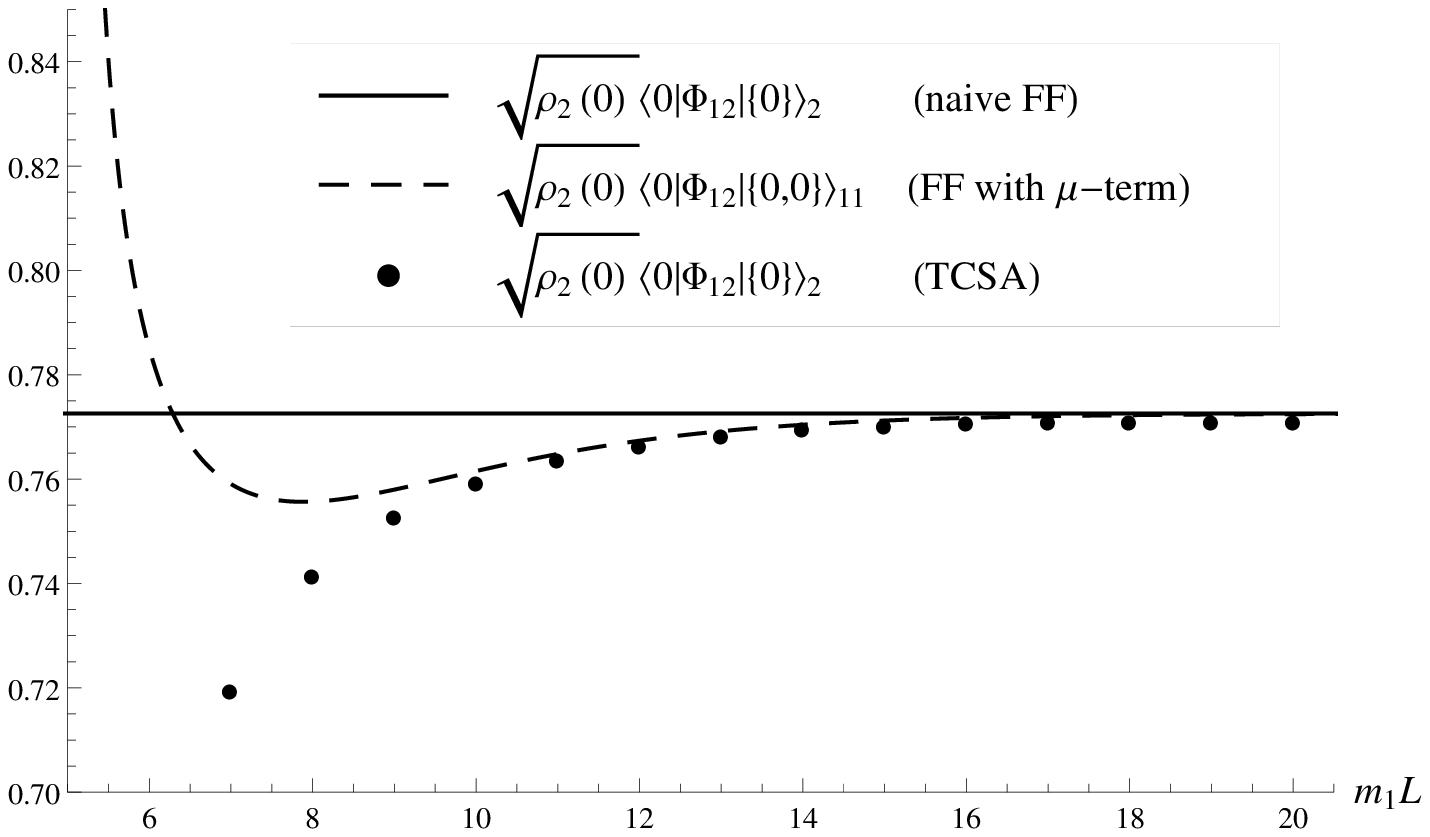}~\\
 \includegraphics{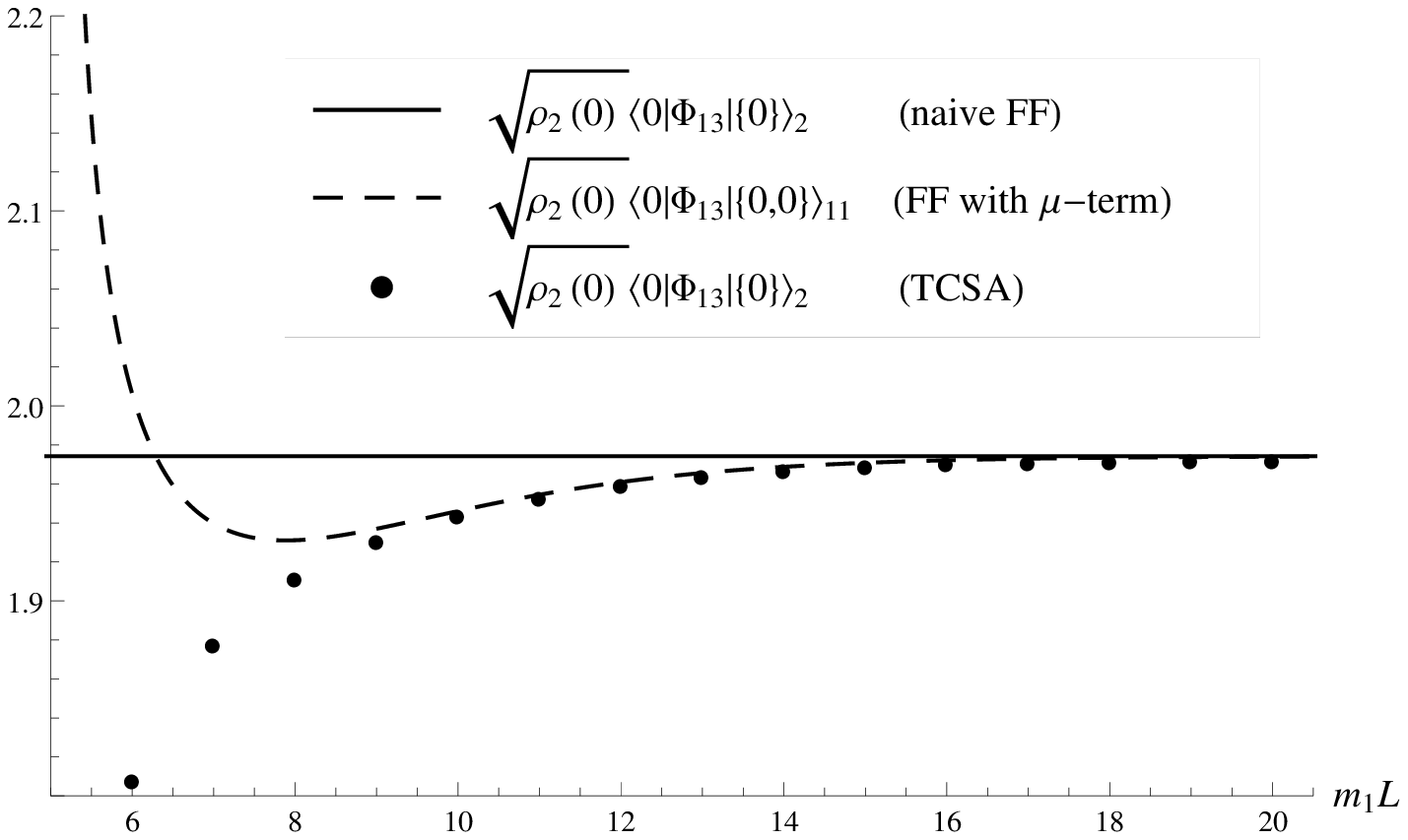} 
\par\end{centering}

\caption{\label{fig:muterm_0_2}Vacuum-$B_{2}$ matrix elements multiplied
by the one-particle density of states for the $B_{2}$ particle for
$\Phi_{1,2}$ and $\Phi_{1,3}$ operators in units of $m_{1}$, where
the solid line is the naive finite volume form factors result, the
dashed line is the correction with the $\mu$-terms and the dots represent
the TCSA data.}
\end{figure}

To end this section, we mention a singularity which is caused by the
inclusion of $\mu$-terms, in a marked contrast to the above two cases
where the $\mu$-terms eliminated singularities of the expression
(\ref{eq:Finite_volume_FF}). Figure \ref{fig:muterm_0_2} shows the
matrix elements multiplied by the one-particle density of states for
the $B_{2}$ particle in the vacuum-$B_{2}$ case, i.e. the expression
\begin{eqnarray}
 &  & \sqrt{\rho_{2}(0)}\langle0|\mathcal{O}|\{0\}\rangle_{2}\nonumber \\
 &  & \mbox{where}\quad\rho_{2}(0)=m_{2}L
\end{eqnarray}
(the extra factor $\sqrt{\rho_{2}(0)}$ is put in to make the large
volume asymptotics of the matrix elements a constant, which makes
the plots easier to interpret). We see that including the $\mu$-term
correction, while improving the agreement with TCSA, also lead to
a singularity at volume $l\sim5.105$. It is plausible that other
exponential corrections would resolve these singularities. The next
class to be taken into account are the so-called $F$-terms, which
would lead to a change in the quantization condition by terms that
vanish exponentially with increasing volume. For the energy levels,
such terms can be found from the excited TBA approach of \cite{Bazhanov:1996aq,Dorey:1996re,Dorey:1997rb};
a general treatment of $F$-terms is given in \cite{Bajnok:2008bm}.
However, for the form factors the theoretical description of these
corrections is not yet developed and is currently under active investigation
\cite{Pozsgay2013}.

\section{Conclusions and outlook \label{sec:Conclusions-and-outlook}}

In this paper we investigated expectation values of local operators
in integrable quantum field theories, in finite volume or, equivalently,
at finite temperature. We considered both numerical and analytical
approaches, and established their consistency, thereby showing that
a combination of these methods is able to give an accurate description
of these expectation values for all scales from small to large volumes,
i.e. from the high-temperature to the the low-temperature regime.
For detailed studies we considered the so-called $T_{2}$ model, which
has a nontrivial primary field besides the perturbing one, and all
of its form factors are known explicitly to any number of particles.

The ultraviolet regime can be accessed using perturbed conformal field
theory, and the non-perturbative truncated conformal space approach
extends its validity significantly and into the infrared regime proper.
When supplemented with renormalization group improvement, it can be
matched to methods based on the infrared date (i.e. scattering theory)
with an astonishing precision in a wide volume/temperature range (indeed,
as shown under \ref{sub:Scaling-Lee-Yang-model-RGTCSA} and \ref{sub:T2-model-RGTCSA},
the RG improvement is mandatory for a proper agreement). The infrared
regime can be accessed using the Leclair-Mussardo series, which for
the perturbing operator is equivalent to the TBA equation. We emphasize
that the matching also works for operators for which the LM series
cannot be derived in the TBA framework, as exemplified by the $\Phi_{1,2}$
operator in the $T_{2}$ model.

In addition to the above results, the paper also contains a derivation
of the cut-off dependence of expectation values in TCSA, which has
already been used before \cite{Konik:2007cb}, but without the formalism
presented; the details are important in order to facilitate the use
of this method in other calculations. Finally, while checking the
accuracy of TCSA and the details of the $T_{2}$ form factors (\ref{eq:b1ffs})
using the methods developed in \cite{Pozsgay:2007kn,Pozsgay:2007gx},
we found interesting cases in which the finite volume form factor
formula, which neglects corrections decaying exponentially with the
volume, predicts singularities that are not observed in TCSA and have
no theoretical reason to exist. It was shown how these singularities
are resolved by including the class of exponential corrections called
$\mu$-terms. However, somewhat surprisingly, this leads to the appearance
of new singularities, which are in turn expected to be cured by higher
exponential corrections. These higher exponential are not yet understood
and it is clear that a systematic description of exponential corrections
to matrix elements in finite volume is necessary to settle all these
issues. Such a description can also have other applications in understanding
the volume/temperature dependence of physical quantities. Work is
in progress in this direction.

\subsection*{Acknowledgments}

GMTW thanks STFC for partial support under grant ST/J002798/1. GT
was partially supported by a Hungarian Academy of Sciences ``Momentum''
grant LP2012-50/2012 and by OTKA grant K81461. We would also like
to thank the referee for comments on the manuscript

\appendix
\makeatletter \renewcommand{\theequation}{\hbox{\normalsize\Alph{section}.\arabic{equation}}} \@addtoreset{equation}{section} \renewcommand{\thefigure}{\hbox{\normalsize\Alph{section}.\arabic{figure}}} \@addtoreset{figure}{section} \renewcommand{\thetable}{\hbox{\normalsize\Alph{section}.\arabic{table}}} \@addtoreset{table}{section} \makeatother

\section{TCSA in the $T_{2}$ model \label{sec:TCSA-in-T2}}

The minimal models $\mathcal{M}_{2,2n+3}$ have the central charge
\cite{Belavin:1984vu} 
\begin{equation}
c=-\frac{2n(5+6n)}{3+2n}
\end{equation}
When perturbed by the relevant operator $\Phi_{1,3}$ with conformal
dimensions 
\begin{equation}
h_{1,3}=\bar{h}_{1,3}=\frac{1-2n}{3+2n}
\end{equation}
for an appropriate choice of the coupling constant they give rise
to the massive quantum field theories $T_{n}$; the $n=1$ case ($T_{1}$)
is the scaling Lee-Yang model. 

For $n=2$ the central charge is 
\begin{equation}
c=-\frac{68}{7}
\end{equation}
and there are two nontrivial primary field $\Phi_{1,2}$ and $\Phi_{1,3}$
with conformal weights $h_{1,2}=\bar{h}_{1,2}=-2/7$ and $h_{1,3}=\bar{h}_{1,3}=-3/7$
and fusion rules 

\begin{align}
\Phi_{1,2}\times\Phi_{1,2} & =1+\Phi_{1,3}\nonumber \\
\Phi_{1,2}\times\Phi_{1,3} & =\Phi_{1,2}+\Phi_{1,3}\nonumber \\
\Phi_{1,3}\times\Phi_{1,3} & =1+\Phi_{1,2}+\Phi_{1,3}\label{eq:t2operules}
\end{align}
where it is helpful to note the identifications $\Phi_{1,2}\equiv\Phi_{1,5}$
and $\Phi_{1,3}\equiv\Phi_{1,4}$.

We consider the theory on a Euclidean space-time cylinder of circumference
$L$ which can be mapped unto the punctured complex plane using 
\begin{equation}
w=\exp\frac{2\pi}{L}(\tau-ix)\qquad,\qquad\bar{w}=\exp\frac{2\pi}{L}(\tau+ix)\label{eq:exponentialmap}
\end{equation}
under which primary fields transform as 
\begin{equation}
\Phi(\tau,x)=\left(\frac{2\pi w}{L}\right)^{h}\left(\frac{2\pi\bar{w}}{L}\right)^{\bar{h}}\Phi(w,\bar{w})
\end{equation}
The fields $\Phi_{1,k}$ are normalized as 
\begin{equation}
\langle0|\Phi_{1,k}(w,\bar{w})\Phi_{1,k}(0,0)|0\rangle=\frac{1}{w^{2h_{1,k}}\bar{w}^{2\bar{h}_{1,k}}}\label{eq:lyconfope}
\end{equation}
and the Hilbert space is given by 
\begin{equation}
\mathcal{H}_{T_{2}}=\bigoplus_{h=0,-2/7,-3/7}\mathcal{V}_{h}\otimes\bar{\mathcal{V}}_{h}
\end{equation}
where $\mathcal{V}_{h}$ ($\bar{\mathcal{V}}_{h}$) denotes the irreducible
representation of the left (right) Virasoro algebra with highest weight
$h$.

The Hamiltonian of the $T_{2}$ model takes the following form in
the perturbed conformal field theory framework: 
\begin{equation}
H=H_{0}+\lambda\int_{0}^{L}dx\Phi_{1,3}(0,x)\label{eq:t2pcftham}
\end{equation}
where 
\begin{equation}
H_{0}=\frac{2\pi}{L}\left(L_{0}+\bar{L}_{0}-\frac{c}{12}\right)
\end{equation}
is the conformal Hamiltonian. When $\lambda>0$ the theory above has
two particles in its spectrum with masses $m_{1}$ and 
\begin{equation}
m_{2}=2m_{1}\cos\frac{\pi}{5}\label{eq:t2massratio}
\end{equation}
The mass gap $m_{1}$ can be related to the coupling constant as \cite{Fateev:1993av}
\begin{eqnarray}
\lambda & = & \kappa m_{1}^{20/7}\label{eq:t2massgap}\\
\mbox{where} &  & \kappa=-0.04053795542\dots\nonumber 
\end{eqnarray}
and the bulk energy density is given by 
\begin{equation}
\mathcal{B}=-\frac{m_{1}^{2}}{8\sin\frac{2\pi}{5}}\label{eq:t2bulk}
\end{equation}
Due to translational invariance of the Hamiltonian (\ref{eq:t2pcftham}),
the conformal Hilbert space $\mathcal{H}$ can be split into sectors
characterized by the eigenvalues of the total spatial momentum 
\begin{equation}
P=\frac{2\pi}{L}\left(L_{0}-\bar{L}_{0}\right)
\end{equation}
where the operator $L_{0}-\bar{L}_{0}$ generates spatial translations
and its eigenvalue is called the conformal spin. For a numerical evaluation
of the spectrum, the Hilbert space is truncated by imposing a cut
in the conformal energy. The truncated conformal space corresponding
to a given truncation and fixed value $s$ of the Lorentz spin reads
\begin{equation}
\mathcal{H}_{\mathrm{TCS}}(s,e_{\mathrm{cut}})=\left\{ |\psi\rangle\in\mathcal{H}\:|\;\left(L_{0}-\bar{L}_{0}\right)|\psi\rangle=s|\psi\rangle,\;\left(L_{0}+\bar{L}_{0}-\frac{c}{12}\right)|\psi\rangle=e|\psi\rangle\,:\, e\leq e_{\mathrm{cut}}\right\} 
\end{equation}
On this subspace, the dimensionless Hamiltonian matrix can be written
as 
\begin{equation}
h_{ij}=\frac{2\pi}{l}\left(L_{0}+\bar{L}_{0}-\frac{c}{12}+\frac{\kappa l^{2-2h_{1,3}}}{(2\pi)^{1-2h_{1.3}}}G^{(s)-1}B^{(s)}\right)\label{eq:dimlesstcsaham}
\end{equation}
where energy is measured in units of the particle mass $m_{1}$, $l=m_{1}L$
is the dimensionless volume parameter, 
\begin{equation}
G_{ij}^{(s)}=\langle i|j\rangle\label{eq:Gs}
\end{equation}
is the conformal inner product matrix and 
\begin{equation}
B_{ij}^{(s)}=\left.\langle i|\Phi_{1,3}(w,\bar{w})|j\rangle\right|_{w=\bar{w}=1}\label{eq:Bs}
\end{equation}
is the matrix element of the operator $\Phi$ at the point $w=\bar{w}=1$
on the complex plane between vectors $|i\rangle$, $|j\rangle$ from
$\mathcal{H}_{\mathrm{TCS}}(s,e_{\mathrm{cut}})$. The natural basis
provided by the action of Virasoro generators is not orthonormal and
therefore $G^{(s)-1}$ must be inserted to transform the left vectors
to the dual basis. The Hilbert space and the matrix elements are constructed
using an algorithm developed by Kausch et al. \cite{Kausch:1996vq}.

Diagonalizing the matrix $h_{ij}$ we obtain the energy levels as
functions of the volume, with energy and length measured in units
of $m$. We considered sectors with $s=0,1,2,3$, and the maximum
value of the cutoff $e_{\mathrm{cut}}$ was $30$, in which case the
Hilbert space contains around twelve thousand vectors for even, and
nine thousand vectors for odd values of the conformal spin. Once the
eigenvectors are obtained, the matrix elements of local operators
can be computed using the exponential mapping to evaluate matrix elements;
for details cf. \cite{Pozsgay:2007kn}.

\section{Scattering theory and form factors of the $T_{2}$ model \label{sec:Scattering-theory-of-T2}}

The form factors of $T_{n}$ models were constructed in \cite{Koubek:1994gk}
using the fact that they can be obtained as reductions of sine-Gordon
theory at a particular value of the coupling \cite{Reshetikhin:1989qg}.
In these particular models, only the breather states are retained
in the spectrum. For $T_{2}$ only the first two breathers $B_{1}$
and $B_{2}$ are retained in the spectrum, and their two-particle
$S$ matrices take the form 
\begin{eqnarray}
S_{11}(\theta) & = & \left\{ \frac{2}{5}\right\} _{\theta}\qquad S_{12}(\theta)=\left\{ \frac{1}{5}\right\} _{\theta}\left\{ \frac{3}{5}\right\} _{\theta}\qquad S_{22}(\theta)=\left\{ \frac{2}{5}\right\} _{\theta}^{2}\left\{ \frac{4}{5}\right\} _{\theta}\nonumber \\
\mbox{with} &  & \left\{ x\right\} _{\theta}=\frac{\sinh\theta+i\sin\pi x}{\sinh\theta-i\sin\pi x}\label{eq:t2smats}
\end{eqnarray}
The form factors of primary fields with the fundamental particle can
be obtained from the form factors of exponential operators in sine-Gordon
theory with the first breather. The latter are given by 
\begin{eqnarray}
F_{\underbrace{{\scriptstyle 11\dots1}}_{n}}^{a}(\theta_{1},\dots,\theta_{n}) & = & \left\langle 0\left|\mathrm{e}^{ia\beta\Phi(0)}\right|B_{1}(\theta_{1})\dots B_{1}(\theta_{n})\right\rangle \nonumber \\
 & = & \mathcal{G}_{a}(\xi)\,[a]_{\xi}\,(i\bar{\lambda}(\xi))^{n}\,\prod_{i<j}\frac{f_{\xi}(\theta_{j}-\theta_{i})}{\mathrm{e}^{\theta_{i}}+\mathrm{e}^{\theta_{j}}}\, Q_{a}^{(n)}\left(\mathrm{e}^{\theta_{1}},\dots,\mathrm{e}^{\theta_{n}}\right)\label{eq:b1ffs}
\end{eqnarray}
where the parameter $\xi$ is 
\begin{equation}
\xi=\frac{\beta^{2}}{8\pi-\beta^{2}}
\end{equation}
and 
\begin{eqnarray}
Q_{a}^{(n)}(x_{1},\dots,x_{n}) & = & \det\left\{[a+i-j]_{\xi}\,\sigma_{2i-j}^{(n)}(x_{1},\dots,x_{n})\right\}_{i,j=1,\dots,n-1}\mbox{ if }n\geq2\nonumber \\
Q_{a}^{(1)} & = & 1\quad,\qquad[a]_{\xi}=\frac{\sin\pi\xi a}{\sin\pi\xi}\\
\bar{\lambda}(\xi) & = & 2\cos\frac{\pi\xi}{2}\sqrt{2\sin\frac{\pi\xi}{2}}\exp\left(-\int_{0}^{\pi\xi}\frac{dt}{2\pi}\frac{t}{\sin t}\right)
\end{eqnarray}
with the function 
\begin{eqnarray}
f_{\xi}(\theta) & = & v(i\pi+\theta,-1)v(i\pi+\theta,-\xi)v(i\pi+\theta,1+\xi)\nonumber \\
 &  & v(-i\pi-\theta,-1)v(-i\pi-\theta,-\xi)v(-i\pi-\theta,1+\xi)\nonumber \\
v(\theta,\zeta) & = & \prod_{k=1}^{N}\left(\frac{\theta+i\pi(2k+\zeta)}{\theta+i\pi(2k-\text{\ensuremath{\zeta}})}\right)^{k}\exp\Bigg\{\int_{0}^{\infty}\frac{dt}{t}\Big(-\frac{\zeta}{4\sinh\frac{t}{2}}-\frac{i\text{\ensuremath{\zeta}}\theta}{2\pi\cosh\frac{t}{2}}\nonumber \\
 &  & +\left(N+1-N\mbox{e}^{-2t}\right)\mbox{e}^{-2Nt+\frac{it\theta}{\pi}}\frac{\sinh\zeta t}{2\sinh^{2}t}\Big)\Bigg\}\label{eq:brminff}
\end{eqnarray}
giving the minimal $B_{1}B_{1}$ form factor%
\footnote{The formula for the function $v$ is in fact independent of $N$;
choosing $N$ large extends the width of the strip where the integral
converges and also speeds up convergence.%
}, while $\sigma_{k}^{(n)}$ denotes the elementary symmetric polynomial
of $n$ variables and order $k$ defined by 
\begin{equation}
\prod_{i=1}^{n}(x+x_{i})=\sum_{k=0}^{n}x^{n-k}\sigma_{k}^{(n)}(x_{1},\dots,x_{n})
\end{equation}
and $\mathcal{G}_{a}(\beta)$ is the vacuum expectation value of the
field which is known exactly \cite{Lukyanov:1996jj}. Form factors
of higher breathers can be computed by representing them as bound
states of $B_{1}$ particles; a useful formula for their evaluation
can be found in Appendix A of \cite{Takacs:2009fu}.

The models $T_{n}$ correspond to restriction at the coupling 
\begin{equation}
\xi=\frac{2}{2n+1}
\end{equation}
For the $T_{2}$ model, the restricted spectrum is composed of the
first and the second breathers $B_{1}$ and $B_{2}$. The form factors
of the operators $\Phi_{1,2}$ and $\Phi_{1,3}$ can be obtained as
the cases $a=1$ and $a=2$ from formula (\ref{eq:b1ffs}) \cite{Koubek:1994gk};
however, the vacuum expectation value needs to be replaced by the
exact vacuum expectation value of the minimal model fields, derived
in \cite{Fateev:1997yg}: 
\begin{eqnarray}
\left\langle \Phi_{1,2}\right\rangle  & = & -2.3251365527\dots\times i\, m_{1}^{-4/7}\nonumber \\
\left\langle \Phi_{1,3}\right\rangle  & = & 2.2695506880\dots\times m_{1}^{-6/7}\label{eq:t2vevs}
\end{eqnarray}
A form factor containing $n$ $B_{1}$ particles and $m$ $B_{2}$
particles can be evaluated using a fundamental form factor (\ref{eq:b1ffs})
containing $n+2m$ $B_{1}$ particles; therefore the connected form
factor in (\ref{eq:connectedFF}) can be evaluated from a $2n+4m$-particle
fundamental form factor.

\providecommand{\href}[2]{#2}\begingroup\raggedright\endgroup


\begin{thebibliography}{10}

\bibitem{Leclair:1999ys}
A.~Leclair and G.~Mussardo, ``{Finite temperature correlation functions in
  integrable QFT},''
  \href{http://dx.doi.org/10.1016/S0550-3213(99)00280-1}{{\em Nucl.Phys.}
  {\bfseries B552} (1999) 624--642},
\href{http://arxiv.org/abs/hep-th/9902075}{{\ttfamily arXiv:hep-th/9902075
  [hep-th]}}.

\bibitem{Yang:1968rm}
C.-N. Yang and C.~Yang, ``{Thermodynamics of one-dimensional system of bosons
  with repulsive delta function interaction},''
{\em J.Math.Phys.} {\bfseries 10} (1969) 1115--1122.

\bibitem{Zamolodchikov:1989cf}
A.~Zamolodchikov, ``{Thermodynamic Bethe Ansatz in Relativistic Models. Scaling
  Three State Potts and Lee-Yang Models},''
\href{http://dx.doi.org/10.1016/0550-3213(90)90333-9}{{\em Nucl.Phys.}
  {\bfseries B342} (1990) 695--720}.

\bibitem{Karowski:1978vz}
M.~Karowski and P.~Weisz, ``{Exact Form-Factors in (1+1)-Dimensional Field
  Theoretic Models with Soliton Behavior},''
\href{http://dx.doi.org/10.1016/0550-3213(78)90362-0}{{\em Nucl. Phys.}
  {\bfseries B139} (1978) 455}.

\bibitem{Kirillov:1987jp}
A.~N. Kirillov and F.~A. Smirnov, ``{A representation of the current algebra
  connected with the SU(2) invariant Thirring model},''
\href{http://dx.doi.org/10.1016/0370-2693(87)90908-7}{{\em Phys. Lett.}
  {\bfseries B198} (1987) 506--510}.

\bibitem{Smirnov:1992vz}
F.~A. Smirnov, ``{Form-factors in completely integrable models of quantum field
  theory},''
{\em Adv. Ser. Math. Phys.} {\bfseries 14} (1992) 1--208.

\bibitem{Saleur:1999hq}
H.~Saleur, ``{A comment on finite temperature correlations in integrable
  QFT},'' \href{http://dx.doi.org/10.1016/S0550-3213(99)00665-3}{{\em Nucl.
  Phys.} {\bfseries B567} (2000) 602--610},
\href{http://arxiv.org/abs/hep-th/9909019}{{\ttfamily arXiv:hep-th/9909019}}.

\bibitem{Pozsgay:2007kn}
B.~Pozsgay and G.~Takacs, ``{Form factors in finite volume I: form factor
  bootstrap and truncated conformal space},''
  \href{http://dx.doi.org/10.1016/j.nuclphysb.2007.06.027}{{\em Nucl. Phys.}
  {\bfseries B788} (2008) 167--208},
\href{http://arxiv.org/abs/0706.1445}{{\ttfamily arXiv:0706.1445 [hep-th]}}.

\bibitem{Pozsgay:2007gx}
B.~Pozsgay and G.~Takacs, ``{Form factors in finite volume. II. Disconnected
  terms and finite temperature correlators},''
  \href{http://dx.doi.org/10.1016/j.nuclphysb.2007.07.008}{{\em Nucl.Phys.}
  {\bfseries B788} (2008) 209--251},
\href{http://arxiv.org/abs/0706.3605}{{\ttfamily arXiv:0706.3605 [hep-th]}}.

\bibitem{Pozsgay:2010xd}
B.~Pozsgay, ``{Mean values of local operators in highly excited Bethe
  states},'' \href{http://dx.doi.org/10.1088/1742-5468/2011/01/P01011}{{\em
  J.Stat.Mech.} {\bfseries 1101} (2011) P01011},
\href{http://arxiv.org/abs/1009.4662}{{\ttfamily arXiv:1009.4662 [hep-th]}}.

\bibitem{Kormos:2009zz}
M.~Kormos, A.~Trombettoni, and G.~Mussardo, ``{Expectation Values in the
  Lieb-Liniger Bose Gas},''
  \href{http://dx.doi.org/10.1103/PhysRevLett.103.210404}{{\em Phys.Rev.Lett.}
  {\bfseries 103} (2009) 210404}.
Selected for the December 2009 issue of Virtual Journal of Atomic Quantum
  Fluids, Vol.1, Issue 6.

\bibitem{Kormos:2010rg}
M.~Kormos, G.~Mussardo, and A.~Trombettoni, ``{Local Correlations in the Super
  Tonks-Girardeau Gas},'' {\em Phys.Rev.} {\bfseries A83} (2011) 013617,
\href{http://arxiv.org/abs/1008.4383}{{\ttfamily arXiv:1008.4383
  [cond-mat.quant-gas]}}.

\bibitem{Pozsgay:2011ec}
B.~Pozsgay, ``{Local correlations in the 1D Bose gas from a scaling limit of
  the XXZ chain},''
  \href{http://dx.doi.org/10.1088/1742-5468/2011/11/P11017}{{\em J.Stat.Mech.}
  {\bfseries 1111} (2011) P11017},
\href{http://arxiv.org/abs/1108.6224}{{\ttfamily arXiv:1108.6224
  [cond-mat.stat-mech]}}.

\bibitem{Essler:2009zz}
F.~H.~L. Essler and R.~M. Konik, ``{Finite-temperature dynamical correlations
  in massive integrable quantum field theories},''
  \href{http://dx.doi.org/10.1088/1742-5468/2009/09/P09018}{{\em J. Stat.
  Mech.} {\bfseries 0909} (2009) P09018},
\href{http://arxiv.org/abs/0907.0779}{{\ttfamily arXiv:0907.0779
  [cond-mat.str-el]}}.

\bibitem{Essler:2007jp}
F.~H.~L. Essler and R.~M. Konik, ``{Finite-temperature lineshapes in gapped
  quantum spin chains},''
  \href{http://dx.doi.org/10.1103/PhysRevB.78.100403}{{\em Phys. Rev.}
  {\bfseries B78} (2008) 100403},
\href{http://arxiv.org/abs/0711.2524}{{\ttfamily arXiv:0711.2524
  [cond-mat.str-el]}}.

\bibitem{Pozsgay:2010cr}
B.~Pozsgay and G.~Takacs, ``{Form factor expansion for thermal correlators},''
  \href{http://dx.doi.org/10.1088/1742-5468/2010/11/P11012}{{\em J.Stat.Mech.}
  {\bfseries 1011} (2010) P11012},
\href{http://arxiv.org/abs/1008.3810}{{\ttfamily arXiv:1008.3810 [hep-th]}}.

\bibitem{Szecsenyi:2012jq}
I.~M. Szecsenyi and G.~Takacs, ``{Spectral expansion for finite temperature
  two-point functions and clustering},''
  \href{http://dx.doi.org/10.1088/1742-5468/2012/12/P12002}{{\em J.Stat.Mech.}
  {\bfseries 1212} (2012) P12002},
\href{http://arxiv.org/abs/1210.0331}{{\ttfamily arXiv:1210.0331 [hep-th]}}.

\bibitem{Yurov:1989yu}
V.~P. Yurov and A.~B. Zamolodchikov, ``{Truncated conformal space approach to
  scaling Lee-Yang model},''
\href{http://dx.doi.org/10.1142/S0217751X9000218X}{{\em Int. J. Mod. Phys.}
  {\bfseries A5} (1990) 3221--3246}.

\bibitem{Feverati:2006ni}
G.~Feverati, K.~Graham, P.~A. Pearce, G.~Z. T\'oth, and G.~M.~T. Watts, ``A
  renormalization group for the truncated conformal space approach,''
  \href{http://dx.doi.org/10.1088/1742-5468/2008/03/P03011}{{\em J. Stat.
  Mech.} {\bfseries 2008} no.~03, (2008) P03011},
\href{http://arxiv.org/abs/hep-th/0612203}{{\ttfamily arXiv:hep-th/0612203
  [hep-th]}}.

\bibitem{Konik:2007cb}
R.~M. Konik and Y.~Adamov, ``{A Numerical Renormalization Group for Continuum
  One-Dimensional Systems},''
  \href{http://dx.doi.org/10.1103/PhysRevLett.98.147205}{{\em Phys. Rev. Lett.}
  {\bfseries 98} (2007) 147205},
  \href{http://arxiv.org/abs/cond-mat/0701605}{{\ttfamily
  arXiv:cond-mat/0701605 [cond-mat.str-el]}}.

\bibitem{Giokas:2011ix}
P.~Giokas and G.~Watts, ``{The renormalisation group for the truncated
  conformal space approach on the cylinder},''
\href{http://arxiv.org/abs/1106.2448}{{\ttfamily arXiv:1106.2448 [hep-th]}}.

\bibitem{Pozsgay:2008bf}
B.~Pozsgay, ``{Luscher's mu-term and finite volume bootstrap principle for
  scattering states and form factors},''
  \href{http://dx.doi.org/10.1016/j.nuclphysb.2008.04.021}{{\em Nucl.Phys.}
  {\bfseries B802} (2008) 435--457},
\href{http://arxiv.org/abs/0803.4445}{{\ttfamily arXiv:0803.4445 [hep-th]}}.

\bibitem{Palmai:2012kf}
T.~Palmai and G.~Takacs, ``{Diagonal multi-soliton matrix elements in finite
  volume},'' {\em Phys. Rev. D} {\bfseries 87} (2013) 045010,
\href{http://arxiv.org/abs/1209.6034}{{\ttfamily arXiv:1209.6034 [hep-th]}}.

\bibitem{Dotsenko:1984nm}
V.~Dotsenko and V.~Fateev, ``{Conformal Algebra and Multipoint Correlation
  Functions in Two-Dimensional Statistical Models},''
\href{http://dx.doi.org/10.1016/0550-3213(84)90269-4}{{\em Nucl.Phys.}
  {\bfseries B240} (1984) 312}.

\bibitem{Dotsenko:1984ad}
V.~Dotsenko and V.~Fateev, ``{Four Point Correlation Functions and the Operator
  Algebra in the Two-Dimensional Conformal Invariant Theories with the Central
  Charge c \&amp;lt; 1},''
\href{http://dx.doi.org/10.1016/S0550-3213(85)80004-3}{{\em Nucl.Phys.}
  {\bfseries B251} (1985) 691}.

\bibitem{Dotsenko:1985hi}
V.~Dotsenko and V.~Fateev, ``{Operator Algebra of Two-Dimensional Conformal
  Theories with Central Charge C \&amp;lt;= 1},''
\href{http://dx.doi.org/10.1016/0370-2693(85)90366-1}{{\em Phys.Lett.}
  {\bfseries B154} (1985) 291--295}.

\bibitem{Zamolodchikov:1989fp}
A.~Zamolodchikov, ``{Integrals of Motion and S Matrix of the (Scaled) T=T(c)
  Ising Model with Magnetic Field},''
\href{http://dx.doi.org/10.1142/S0217751X8900176X}{{\em Int.J.Mod.Phys.}
  {\bfseries A4} (1989) 4235}.

\bibitem{Feverati:1998va}
G.~Feverati, F.~Ravanini, and G.~Takacs, ``{Truncated conformal space at c = 1,
  nonlinear integral equation and quantization rules for multi - soliton
  states},'' \href{http://dx.doi.org/10.1016/S0370-2693(98)00543-7}{{\em
  Phys.Lett.} {\bfseries B430} (1998) 264--273},
\href{http://arxiv.org/abs/hep-th/9803104}{{\ttfamily arXiv:hep-th/9803104
  [hep-th]}}.

\bibitem{Cardy1989}
J.~L. Cardy and G.~Mussardo, ``{S Matrix of the Yang-Lee Edge Singularity in
  Two-Dimensions},''
\href{http://dx.doi.org/10.1016/0370-2693(89)90818-6}{{\em Phys.Lett.}
  {\bfseries B225} (1989) 275}.

\bibitem{Fateev:1997yg}
V.~Fateev, S.~L. Lukyanov, A.~B. Zamolodchikov, and A.~B. Zamolodchikov,
  ``{Expectation values of local fields in Bullough-Dodd model and integrable
  perturbed conformal field theories},''
  \href{http://dx.doi.org/10.1016/S0550-3213(98)00002-9}{{\em Nucl.Phys.}
  {\bfseries B516} (1998) 652--674},
\href{http://arxiv.org/abs/hep-th/9709034}{{\ttfamily arXiv:hep-th/9709034
  [hep-th]}}.

\bibitem{Dorey:2000eh}
P.~Dorey, M.~Pillin, R.~Tateo, and G.~Watts, ``{One point functions in
  perturbed boundary conformal field theories},''
  \href{http://dx.doi.org/10.1016/S0550-3213(00)00622-2}{{\em Nucl.Phys.}
  {\bfseries B594} (2001) 625--659},
\href{http://arxiv.org/abs/hep-th/0007077}{{\ttfamily arXiv:hep-th/0007077
  [hep-th]}}.

\bibitem{Kausch:1996vq}
H.~Kausch, G.~Takacs, and G.~Watts, ``{On the relation between Phi(1,2) and
  Phi(1,5) perturbed minimal models},''
  \href{http://dx.doi.org/10.1016/S0550-3213(97)00056-4}{{\em Nucl.Phys.}
  {\bfseries B489} (1997) 557--579},
\href{http://arxiv.org/abs/hep-th/9605104}{{\ttfamily arXiv:hep-th/9605104
  [hep-th]}}.

\bibitem{Feher:2011aa}
G.~Feher and G.~Takacs, ``{Sine-Gordon form factors in finite volume},''
  \href{http://dx.doi.org/10.1016/j.nuclphysb.2011.06.020}{{\em Nucl.Phys.}
  {\bfseries B852} (2011) 441--467},
\href{http://arxiv.org/abs/1106.1901}{{\ttfamily arXiv:1106.1901 [hep-th]}}.

\bibitem{Takacs:2011nb}
G.~Takacs, ``{Determining matrix elements and resonance widths from finite
  volume: the dangerous $\mu$-terms},''
  \href{http://dx.doi.org/10.1007/JHEP11(2011)113}{{\em JHEP} {\bfseries 1111}
  (2011) 113},
\href{http://arxiv.org/abs/1110.2181}{{\ttfamily arXiv:1110.2181 [hep-th]}}.

\bibitem{Bazhanov:1996aq}
V.~V. Bazhanov, S.~L. Lukyanov, and A.~B. Zamolodchikov, ``{Integrable quantum
  field theories in finite volume: Excited state energies},''
  \href{http://dx.doi.org/10.1016/S0550-3213(97)00022-9}{{\em Nucl.Phys.}
  {\bfseries B489} (1997) 487--531},
\href{http://arxiv.org/abs/hep-th/9607099}{{\ttfamily arXiv:hep-th/9607099
  [hep-th]}}.

\bibitem{Dorey:1996re}
P.~Dorey and R.~Tateo, ``{Excited states by analytic continuation of TBA
  equations},'' \href{http://dx.doi.org/10.1016/S0550-3213(96)00516-0}{{\em
  Nucl.Phys.} {\bfseries B482} (1996) 639--659},
\href{http://arxiv.org/abs/hep-th/9607167}{{\ttfamily arXiv:hep-th/9607167
  [hep-th]}}.

\bibitem{Dorey:1997rb}
P.~Dorey and R.~Tateo, ``{Excited states in some simple perturbed conformal
  field theories},''
  \href{http://dx.doi.org/10.1016/S0550-3213(97)00838-9}{{\em Nucl.Phys.}
  {\bfseries B515} (1998) 575--623},
\href{http://arxiv.org/abs/hep-th/9706140}{{\ttfamily arXiv:hep-th/9706140
  [hep-th]}}.

\bibitem{Bajnok:2008bm}
Z.~Bajnok and R.~A. Janik, ``{Four-loop perturbative Konishi from strings and
  finite size effects for multiparticle states},''
  \href{http://dx.doi.org/10.1016/j.nuclphysb.2008.08.020}{{\em Nucl.Phys.}
  {\bfseries B807} (2009) 625--650},
\href{http://arxiv.org/abs/0807.0399}{{\ttfamily arXiv:0807.0399 [hep-th]}}.

\bibitem{Pozsgay2013}
B.~Pozsgay, ``{Form factor approach to diagonal finite volume matrix elements
  in Integrable QFT},''
\href{http://arxiv.org/abs/1305.3373}{{\ttfamily arXiv:1305.3373 [hep-th]}}.

\bibitem{Belavin:1984vu}
A.~Belavin, A.~M. Polyakov, and A.~Zamolodchikov, ``{Infinite Conformal
  Symmetry in Two-Dimensional Quantum Field Theory},''
\href{http://dx.doi.org/10.1016/0550-3213(84)90052-X}{{\em Nucl.Phys.}
  {\bfseries B241} (1984) 333--380}.

\bibitem{Fateev:1993av}
V.~Fateev, ``{The Exact relations between the coupling constants and the masses
  of particles for the integrable perturbed conformal field theories},''
\href{http://dx.doi.org/10.1016/0370-2693(94)00078-6}{{\em Phys.Lett.}
  {\bfseries B324} (1994) 45--51}.

\bibitem{Koubek:1994gk}
A.~Koubek, ``{Form-factor bootstrap and the operator content of perturbed
  minimal models},'' \href{http://dx.doi.org/10.1016/0550-3213(94)90368-9}{{\em
  Nucl.Phys.} {\bfseries B428} (1994) 655--680},
\href{http://arxiv.org/abs/hep-th/9405014}{{\ttfamily arXiv:hep-th/9405014
  [hep-th]}}.

\bibitem{Reshetikhin:1989qg}
N.~Reshetikhin and F.~Smirnov, ``{Hidden Quantum Group Symmetry and Integrable
  Perturbations of Conformal Field Theories},''
\href{http://dx.doi.org/10.1007/BF02097683}{{\em Commun.Math.Phys.} {\bfseries
  131} (1990) 157--178}.

\bibitem{Lukyanov:1996jj}
S.~L. Lukyanov and A.~B. Zamolodchikov, ``{Exact expectation values of local
  fields in quantum sine-Gordon model},''
  \href{http://dx.doi.org/10.1016/S0550-3213(97)00123-5}{{\em Nucl.Phys.}
  {\bfseries B493} (1997) 571--587},
\href{http://arxiv.org/abs/hep-th/9611238}{{\ttfamily arXiv:hep-th/9611238
  [hep-th]}}.

\bibitem{Takacs:2009fu}
G.~Takacs, ``{Form factor perturbation theory from finite volume},''
  \href{http://dx.doi.org/10.1016/j.nuclphysb.2009.10.001}{{\em Nucl. Phys.}
  {\bfseries B825} (2010) 466--481},
\href{http://arxiv.org/abs/0907.2109}{{\ttfamily arXiv:0907.2109 [hep-th]}}.

\end{thebibliography}
\end{document}